\documentclass{article}
\usepackage{graphicx}
\usepackage{amsmath}
\usepackage{amssymb}
\usepackage{epubtk}
\usepackage{mdwlist} 

\showlistoftablesfalse

\def\slambda{{\ensuremath{\lower0.3ex\hbox{\char"16}}\kern-0.6em\lambda}}

\newcommand{\be}{\begin{equation}}
\newcommand{\dst}{\displaystyle}
\newcommand{\ee}{\end{equation}}

\newcommand{\F}{{\mathcal{F}}}
\newcommand{\B}{{\mathcal{B}}}
\newcommand{\Fo}{{\mathcal{F}_0}}
\newcommand{\Fs}{{\mathcal{F}_\text{s}}}

\newcommand{\gw}{gra\-vi\-ta\-tio\-nal-wave }

\newcommand{\md}{\mathrm{d}}
\newcommand{\mE}{\mathrm{E}}

\newcommand{\To}{{T_{\mathrm{o}}}}

\newcommand{\mathbold}[1]{\mbox{\boldmath $#1$}}
\newcommand{\mathboldsmall}[1]{\mbox{\scriptsize \boldmath $#1$}}

\newcommand{\bzeta}{\boldsymbol{\zeta}}
\newcommand{\btheta}{\boldsymbol{\theta}}
\newcommand{\bxi}{\boldsymbol{\xi}}

\newcommand{\nc}{{N_\text{cells}}}

\DeclareMathOperator{\tr}{Tr}

\begin{document}

\title{Gravitational-Wave Data Analysis. \\
       Formalism and Sample Applications: The Gaussian Case}

\author{
\epubtkAuthorData{Piotr Jaranowski}
                 {Faculty of Physics \\
                  University of Bia{\l}ystok \\
                  Lipowa 41 \\
                  15--424 Bia{\l}ystok, Poland}
                 {pio@alpha.uwb.edu.pl}
                 {}
\and
\epubtkAuthorData{Andrzej Kr\'olak}
                 {Institute of Mathematics \\
                  Polish Academy of Sciences \\
                  \'Sniadeckich 8 \\
                  00--956 Warsaw, Poland}
                 {krolak@impan.gov.pl}
                 {}
}

\date{}
\maketitle


\begin{abstract}
  The article reviews the statistical theory of signal detection in
  application to analysis of deterministic gravitational-wave signals
  in the noise of a detector. Statistical foundations for the theory
  of signal detection and parameter estimation are presented. Several
  tools needed for both theoretical evaluation of the optimal data
  analysis methods and for their practical implementation are
  introduced. They include optimal signal-to-noise ratio, Fisher
  matrix, false alarm and detection probabilities, $\F$-statistic,
  template placement, and fitting factor. These tools apply to the
  case of signals buried in a stationary and Gaussian
  noise. Algorithms to efficiently implement the optimal data analysis
  techniques are discussed. Formulas are given for a general
  gravitational-wave signal that includes as special cases most of the
  deterministic signals of interest.
\end{abstract}

\epubtkKeywords{gravitational waves, signal detection, parameter
  estimation}

\newpage


\epubtkUpdate
    [Id=B,
     ApprovedBy=subjecteditor,
     AcceptDate={14 February 2012},
     PublishDate={9 March 2012},
     Type=major]{%
Material of the previous version of the review was partially reorganized
and updated, 46 new references were added.

1. Section~\ref{sec:response} was rewritten and extended, several new
references were added.

2. Some parts of the former Section~4 were moved to the
present Section~\ref{sec:statistical-theory}, which is now a brief
general introduction to the statistical theory of signal detection and of
estimation of signals parameters. Some new references were added.

3. The present Section~\ref{sec:Fisher-matrix-Gaussian-case} is a
partially rewritten (using some new, more convenient notation) and
extended version of the former Sections 4.3\,--\,4.9. The
gravitational-wave signal considered here was generalized from a
4-amplitude-parameter case to an \textit{n}-amplitude-parameter case,
where \textit{n} is arbitrary. New Section~\ref{sec:targeted-searches}
about targeted searches was added, and new
Section~\ref{sec:covering-problem} on the covering problem was created
with references to constructions of various grids of templates for
searches of continuous gravitational waves.

4. The present Section~\ref{sec:network-of-detectors} is an expanded
version of the former Section~4.10 with addition of several recent references.

5. The present Section~\ref{sec:non} is an expanded version of the
former Section~4.11 with new discussion of optimal filtering for
non-stationary data and description of a test (Grubbs' test) to detect
outliers in data.
}


\section{Introduction}
\label{section:introduction}

In this review we consider the problem of detection of
\emph{deterministic} gravitational-wave signals in the noise of a
detector and the question of estimation of their parameters. The
examples of deterministic signals are gravitational waves from
rotating neutron stars, coalescing compact binaries, and supernova
explosions. The case of detection of stochastic gravitational-wave
signals in the noise of a detector is reviewed in~\cite{allen-97}.
A very powerful method to detect a signal in noise that is optimal
by several criteria consists of correlating the data with the
template that is matched to the expected signal. This
\emph{matched-filtering} technique is a special case of the
\emph{maximum likelihood} detection method. In this review we
describe the theoretical foundation of the method and we show
how it can be applied to the case of a very general deterministic
gravitational-wave signal buried in a \emph{stationary} and
\emph{Gaussian} noise.

Early gravitational-wave data analysis was concerned with the
detection of bursts originating from supernova
explosions~\cite{weber-69}. It involved analysis of the coincidences
among the detectors~\cite{kafka-77}. With the growing interest in
laser interferometric gravitational-wave detectors that are broadband
it was realized that sources other than supernovae can also be
detectable~\cite{thorne-87} and that they can provide a wealth of
astrophysical information~\cite{schutz-86, krolak-87}.
For example, the analytic form of the gravitational-wave signal
produced during the inspiral phase of a compact binary coalescence
is known in terms of a few parameters to a good approximation
(see, e.g., \cite{blanchet-06} and Section~2.4 of~\cite{JaranowskiKrolak2009}).
Consequently one can detect such a signal by
correlating the data with the predicted waveform (often called the template)
and maximizing the correlation with respect to the parameters of the waveform.
Using this method one can pick up a weak signal from the
noise by building a large signal-to-noise ratio over a wide
bandwidth of the detector~\cite{thorne-87}. This observation has
led to rapid development of the theory of gravitational-wave
data analysis. It became clear that the detectability of sources is
determined by optimal signal-to-noise ratio,
which is the power spectrum of the signal divided by the power
spectrum of the noise integrated over the bandwidth of the
detector.

An important landmark was a workshop entitled
\emph{Gravitational Wave Data Analysis} held in Dyffryn House and
Gardens, St.~Nicholas near Cardiff, in July 1987~\cite{schutz-89}.
The meeting acquainted physicists interested in analyzing
gravitational-wave data with the basics of the statistical theory
of signal detection and its application to detection of
gravitational-wave sources. As a result of subsequent studies, the
Fisher information matrix was introduced to the theory of the
analysis of gravitational-wave data~\cite{finn-93, krolak-93}. The
diagonal elements of the Fisher matrix give lower bounds on the
variances of the estimators of the parameters of the signal and
can be used to assess the quality of astrophysical information
that can be obtained from detections of gravitational-wave
signals~\cite{cutler-94, krolak-95, bala-98}. It was also realized
that the application of matched-filtering to some sources, notably to
continuous sources originating from neutron stars, will require
extraordinary large computing resources. This gave a further
stimulus to the development of optimal and efficient algorithms
and data analysis methods~\cite{schutz-91}.

A very important development was the work by
Cutler et~al.~\cite{cutler-93} where it was realized that for the case
of coalescing binaries matched filtering was sensitive to very small
post-Newtonian effects of the waveform. Thus, these effects can be
detected. This leads to a much better verification of Einstein's
theory of relativity and provides a wealth of astrophysical
information that would make a laser interferometric gravitational-wave
detector a true astronomical observatory complementary to those
utilizing the electromagnetic spectrum. As further development of the
theory, methods were introduced to calculate the quality of suboptimal
filters~\cite{apostolatos-95}, to calculate the number of templates required to
do a search using matched-filtering~\cite{owen-96}, to determine the
accuracy of templates required~\cite{brady-98}, and to calculate the
false alarm probability and thresholds~\cite{jaranowski-98}. An
important point is the reduction of the number of parameters that one
needs to search for in order to detect a signal. Namely estimators of
a certain type of parameters, called \emph{extrinsic parameters}, can
be found in a closed analytic form and consequently eliminated from
the search. Thus, a computationally-intensive search need only be performed
over a reduced set of \emph{intrinsic parameters}~\cite{krolak-93,
  jaranowski-98, krolak-04}.

Techniques reviewed in this paper have been used in the data
analysis of prototypes of gravita\-tio\-nal-wave
detectors~\cite{niebauer-93, nicholson-96, allen-99} and in the data
analysis of gravita\-tio\-nal-wave
detectors currently in operation~\cite{tama-01, astone-03, ligo1-04, ligo2-04, ligo3-04}.


\newpage

\section{Response of a Detector to a Gravitational Wave}
\label{sec:response}


There are two main methods of detecting gravitational waves currently in use. One method
is to measure changes induced by gravitational waves on the
distances between freely-moving test masses using coherent trains
of electromagnetic waves. The other method is to measure the
deformation of large masses at their resonance frequencies induced
by gravitational waves.
The first idea is realized in laser interferometric detectors
(both Earth-based~\cite{RowanHough2011, Vinet2009, FreiseStrain2010}
and space-borne~\cite{lisa-98, TintoDhurandhar2005} antennas)
and Doppler tracking experiments~\cite{Armstrong2006},
whereas the second idea is implemented in resonant mass detectors
(see, e.g., \cite{astone-93}).

\subsection{Doppler shift between freely falling particles}
\label{subsection:Doppler_shift}

We start by describing the change of a photon's frequency caused
by a passing gravitational wave and registered by particles
(representing different parts of a gravitational-wave detector)
freely falling in the field of the gravitational wave.
The detailed derivation of the formulae we show here
can be found in Chapter~5 of~\cite{JaranowskiKrolak2009}
(see also~\cite{estabrook-75, armstrong-99, rubbo-03}).
An equivalent derivation of the response of test masses
to gravitational waves in the local Lorentz gauge
(without making use of the long-wavelength approximation)
is given in~\cite{Rakhmanov2005}.

We employ here the \emph{transverse} \emph{traceless} (TT)
coordinate system (more about the TT gauge can be found, e.g.,
in Section~35.4 of~\cite{MisnerThorneWheeler1973}
or in Section~1.3 of~\cite{JaranowskiKrolak2009}).
A spacetime metric describing a \emph{plane} gravitational wave
traveling in the $+z$ direction of the TT coordinate system
(with coordinates $x^0\equiv ct$, $x^1\equiv x$, $x^2\equiv y$, $x^3\equiv z$),
is described by the line element
\begin{equation}
\label{r2.1}
\md s^2 = -c^2 \md t^2 + \bigg(1+h_+\Big(t-\frac{z}{c}\Big)\bigg)\md x^2
+ \bigg(1-h_+\Big(t-\frac{z}{c}\Big)\bigg)\md y^2
 + 2\,h_\times\Big(t-\frac{z}{c}\Big) \md x \, \md y + \md z^2,
\end{equation}
where $h_+$ and $h_\times$ are the two independent polarizations of the wave.
We assume that the wave is \emph{weak}, i.e., for any instant of time $t$,
\be
|h_+(t)| \ll 1, \quad |h_\times(t)| \ll 1.
\ee
We will neglect all terms of order $h^2$ or higher.
The form of the line element~\eqref{r2.1} implies
that the functions $h_+(t)$ and $h_\times(t)$
describe the wave-induced perturbation of the flat Minkowski metric
\emph{at the origin} of the TT coordinate system (where $x=y=z=0$).
It is convenient to introduce the three-dimensional matrix
of the spatial metric perturbation produced by the gravitational wave
(at the coordinate system's origin),
\be
\label{r2.2}
\mathsf{H}(t) := \left(
\begin{array}{ccc}
h_+(t) & h_\times(t) & 0 \\
h_\times(t) & -h_+(t) & 0 \\
0 & 0 & 0 \\
\end{array} \right).
\ee

Let two particles freely fall in the field~\eqref{r2.1} of the gravitational wave,
and let their spatial coordinates remain \emph{constant}, so the particles'
world lines are described by equations
\be
\label{r2.3}
t(\tau_a)=\tau_a, \quad x(\tau_a)=x_a, \quad y(\tau_a)=y_a, \quad z(\tau_a)=z_a,
\quad a=1,2,
\ee
where $(x_a,y_a,z_a)$ are spatial coordinates of the $a$th particle
and $\tau_a$ is its proper time.
These two particles measure, in their proper reference frames,
the frequency of the \emph{same} photon traveling
along a null geodesic $x^\alpha=x^\alpha(\lambda)$,
where $\lambda$ is some affine parameter.
The coordinate time, at which the photon's frequency is measured by the $a$th particle,
is equal to $t_a$ ($a=1,2$); we assume that $t_2>t_1$.
Let us introduce the coordinate time duration $t_{12}$ of the photon's trip
and the Euclidean coordinate distance $L_{12}$ between the particles:
\be
\label{r2.4}
t_{12} := t_2 - t_1,
\quad
L_{12} := \sqrt{(x_2-x_1)^2+(y_2-y_1)^2+(z_2-z_1)^2}.
\ee
Let us also introduce the 3-vector $\mathbf{n}$ of unit Euclidean length
directed along the line connecting the two particles.
We arrange the components of this vector into the column $3\times1$ matrix $\mathsf{n}$
(thus, we distinguish here the 3-vector $\mathbf{n}$ from its components
being the elements of the matrix $\mathsf{n}$; the \textit{same} 3-vector
can be decomposed into components in \textit{different} spatial coordinate systems):
\be
\label{r2.5}
\mathsf{n} := (\cos\alpha,\cos\beta,\cos\gamma)^\mathsf{T}
= \begin{pmatrix}
\cos\alpha \\
\cos\beta  \\
\cos\gamma \end{pmatrix},
\ee
where the superscript $\mathsf{T}$ denotes \textit{matrix transposition}.
If one neglects the spacetime curvature caused by the gravitational wave,
then $\alpha$, $\beta$, and $\gamma$ are the angles
between the path of the photon in the 3-space
and the coordinate axis $x$, $y$, or $z$, respectively
(obviously, $\alpha,\beta,\gamma\in\langle0;\pi\rangle$
and $\cos^2\alpha+\cos^2\beta+\cos^2\gamma=1$).
Let us denote the value of the frequency
registered by the $a$th particle by $\nu_a$ ($a=1,2$)
and let us finally define the relative change of the photon's frequencies,
\be
\label{r2.06}
y_{12} := \frac{\nu_2}{\nu_1} - 1.
\ee
Then, it can be shown
(see Chapter~5 of~\cite{JaranowskiKrolak2009} for details)
that the frequency ratio $y_{12}$ can be written
[making use of the quantities introduced
in Eqs.~\eqref{r2.2} and \eqref{r2.4}\,--\,\eqref{r2.5}]
as follows (the dot means here \emph{matrix multiplication}):
\be
\label{r2.07}
y_{12} = \frac{1}{2(1-\cos\gamma)}\,
\mathsf{n}^\mathsf{T} \cdot \bigg(
  \mathsf{H}\Big(t_1-\frac{z_1}{c}\Big)
- \mathsf{H}\Big(t_1-\frac{z_1}{c}+(1-\cos\gamma)\frac{L_{12}}{c}\Big)
\bigg) \cdot \mathsf{n}
+ {\cal O}\big(h^2\big).
\ee

It is convenient to introduce the unit 3-vector $\mathbf{k}$
directed from the origin of the coordinate system
to the source of the gravitational wave.
In the coordinate system adopted by us the wave is traveling in the $+z$ direction.
Therefore, the components of the 3-vector $\mathbf{k}$,
arranged into the column matrix $\mathsf{k}$, are
\be
\label{r2.08}
\mathsf{k} = (0,0,-1)^{\mathsf{T}}.
\ee
The positions of the particles with respect to the origin of the coordinate system
we describe by the 3-vectors $\mathbf{x}_a$ ($a=1,2$),
the components of which we put into the column matrices $\mathsf{x}_a$:
\be
\label{r2.09}
\mathsf{x}_a = (x_a,y_a,z_a)^{\mathsf{T}}, \quad a=1,2.
\ee
Making use of Eqs.~\eqref{r2.08}\,--\,\eqref{r2.09}
we rewrite the basic formula~\eqref{r2.07} in the following form
\be
\label{r2.10}
\dst y_{12} = \frac{\dst \mathsf{n}^\mathsf{T} \! \cdot \bigg(
{\mathsf{H}}\Big(t_1+\frac{\mathsf{k}^\mathsf{T}\cdot\mathsf{x}_1}{c}\Big)
- {\mathsf{H}}\Big(t_1+\frac{L_{12}}{c}+\frac{\mathsf{k}^\mathsf{T}\cdot\mathsf{x}_2}{c}\Big)
\bigg) \cdot \mathsf{n}}
{2(1+\mathsf{k}^\mathsf{T}\cdot\mathsf{n})}
+ {\cal O}\big(h^2\big).
\ee

\begin{figure}[htbp]
\begin{center}
\includegraphics[scale=0.75]{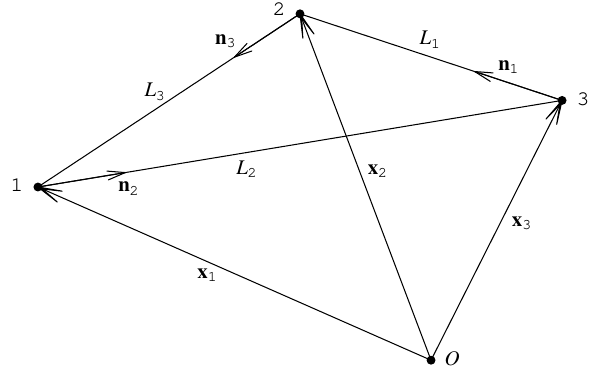}
\caption{Schematic configuration of three freely-falling particles
as a detector of gravitational waves.
The particles are labelled 1, 2, and 3,
their positions with respect to the origin O of the coordinate system
are given by 3-vectors ${\mathbf{x}}_a$ $(a=1,2,3)$.
The Euclidean separations between the particles are denoted by $L_a$,
where the index $a$ corresponds to the opposite particle.
The unit 3-vectors ${\mathbf{n}}_a$ point between pairs of particles,
with the orientation indicated.}
\label{fig:3particles}
\end{center}
\end{figure}

To obtain the response for all currently working and planned detectors
it is enough to consider a configuration of three particles
shown in Figure~\ref{fig:3particles}.
Two particles model a Doppler tracking experiment,
where one particle is the Earth and the other is a distant spacecraft.
Three particles model a ground-based laser interferometer,
where the masses are suspended from seismically-isolated supports
or a space-borne interferometer,
where the three test masses are shielded in satellites
driven by drag-free control systems.
In Figure~\ref{fig:3particles} we have introduced the following notation:
O denotes the origin of the TT coordinate system related to the
passing gravitational wave, ${\mathbf{x}}_a$ ($a=1,2,3$) are 3-vectors
joining O and the particles, ${\mathbf{n}}_a$ and $L_a$ ($a=1,2,3$)
are, respectively, 3-vectors of unit Euclidean length along the lines
joining the particles and the coordinate Euclidean distances between the
particles, where $a$ is the label of the opposite particle.
We still assume that the spatial coordinates of the particles
do not change in time.

Let us denote by $\nu_0$ the frequency of the coherent beam used in the detector
(laser light in the case of an interferometer
and radio waves in the case of Doppler tracking).
Let the particle 1 emit the photon with frequency $\nu_0$ at the moment $t_0$ towards the particle 2,
which registers the photon with frequency $\nu'$ at the moment $t'=t_0+L_3/c+{\cal{O}}(h)$.
The photon is immediately transponded (without change of frequency) back to the particle 1,
which registers the photon with frequency $\nu$ at the moment $t=t_0+2L_3/c+{\cal{O}}(h)$.
We express the relative changes of the photon's frequency
$y_{12}:=(\nu'-\nu_0)/\nu_0$ and $y_{21}:=(\nu-\nu')/\nu'$
as functions of the \textit{instant of time $t$}.
Making use of Eq.~\eqref{r2.10} we obtain
\begin{subequations}
\label{r2.11}
\begin{eqnarray}
y_{12}(t) &=& \frac{1}{2(1-\mathsf{k}^{\mathsf{T}}\cdot\mathsf{n}_3)}
\mathsf{n}_3^\mathsf{T}\cdot\bigg(
{\mathsf{H}}\Big(t-\frac{2L_3}{c}+\frac{\mathsf{k}^{\mathsf{T}}\cdot{\mathsf{x}}_1}{c}\Big)
- {\mathsf{H}}\Big(t-\frac{L_3}{c}+\frac{\mathsf{k}^{\mathsf{T}}\cdot{\mathsf{x}}_2}{c}\Big)
\bigg)\cdot\mathsf{n}_3
+ {\cal{O}}\big(h^2\big),
\\
y_{21}(t) &=& \frac{1}{2(1+\mathsf{k}^{\mathsf{T}}\cdot\mathsf{n}_3)}
\mathsf{n}_3^\mathsf{T}\cdot\bigg(
  {\mathsf{H}}\Big(t-\frac{L_3}{c}+\frac{\mathsf{k}^{\mathsf{T}}\cdot{\mathsf{x}}_2}{c}\Big)
- {\mathsf{H}}\Big(t+\frac{\mathsf{k}^{\mathsf{T}}\cdot{\mathsf{x}}_1}{c}\Big)
\bigg)\cdot\mathsf{n}_3
+ {\cal{O}}\big(h^2\big).
\end{eqnarray}
\end{subequations}
The total frequency shift $y_{121}:=(\nu-\nu_0)/\nu_0$
of the photon during its round trip can be computed
from the one-way frequency shifts $y_{12}$ and $y_{21}$ given above:
\begin{equation}
\label{r2.12}
y_{121} = \frac{\nu}{\nu_0} - 1
= \frac{\nu}{\nu'}\frac{\nu'}{\nu_0} - 1
= (y_{21} + 1)(y_{12} + 1) - 1
= y_{12} + y_{21} + {\cal O}\big(h^2\big).
\end{equation}

\subsection{Long-wavelength approximation}

Let $L$ be the size of the detector and $\slambda:=\lambda/(2\pi)$
be the reduced wavelength of the gravitational wave impinging on the detector.
In the \emph{long-wavelength approximation} the condition $\slambda \gg L$
is fulfilled.
The angular frequency of the wave equals $\omega=c/\slambda$.
Time delays caused by the finite speed of the wave propagating across the detector
are of order $\Delta{t}\sim L/c$, but
\be
\omega\Delta{t} \sim \frac{L}{\slambda} \ll 1,
\ee
so time delays across the detector are much shorter than the period of the gravitational wave
and can be neglected. It means that with a good accuracy the \gw field
can be treated as being uniform (but time-dependent)
in the space region that covers the entire detector.
To detect gravitational waves with some dominant angular frequency $\omega$
one must collect data over time intervals
longer (sometimes much longer) than the \gw period.
This implies that in Eq.~\eqref{r2.07} for the relative frequency shift,
the typical value of the quantity $\bar{t}:=t_1-z_1/c$
will be much larger than the retardation time $\Delta{t}:=L_{12}/c$.
Therefore, we can expand this equation with respect to $\Delta{t}$
and keep terms only linear in $\Delta{t}$. After doing this one obtains
(see Section~5.3 in~\cite{JaranowskiKrolak2009} for more details):
\be
\label{r1.lw1}
y_{12} = -\frac{L_{12}}{2c} \,
\mathsf{n}^\mathsf{T}
\cdot \dot{\mathsf{H}}\left(t_1-\frac{z_1}{c}\right)
\cdot \mathsf{n}
+ {\cal O}\big(h^2,\Delta{t}^2\big),
\ee
where overdot denotes differentiation with respect to time.

For the configuration of particles shown in Figure~\ref{fig:3particles},
the relative frequency shifts $y_{12}$ and $y_{21}$ given by Eqs.~\eqref{r2.11} can be written,
by virtue of the formula~\eqref{r1.lw1}, in the form
\be
\label{r1.lw2}
y_{12}(t) = y_{21}(t) = -\frac{L_3}{2c} \,
\mathsf{n}^\mathsf{T}_3
\cdot \dot{\mathsf{H}}\left(t+\frac{\mathsf{k}^\mathsf{T}\cdot\mathsf{x}_1}{c}\right)
\cdot \mathsf{n}_3
+ {\cal O}\big(h^2,\Delta{t}^2\big),
\ee
so that they are equal to each other up to terms ${\cal O}\big(h^2,\Delta{t}^2\big)$.
The photon's total round-trip frequency shift $y_{121}$
[cf.\ Eq.~\eqref{r2.12}] is thus equal to
\be
\label{r1.lw3}
y_{121}(t) = -\frac{L_3}{c} \,
\mathsf{n}^\mathsf{T}_3
\cdot \dot{\mathsf{H}}\left(t+\frac{\mathsf{k}^\mathsf{T}\cdot\mathsf{x}_1}{c}\right)
\cdot \mathsf{n}_3
+ {\cal O}\big(h^2,\Delta{t}^2\big).
\ee

There are important cases where the long-wavelength approximation is not valid.
These include satellite Doppler tracking measurements
and the space-borne LISA detector
for \gw frequencies larger than a few mHz.

\subsection{Solar-system-based detectors}

Real gravitational-wave detectors do not stay at rest
with respect to the TT coordinate system related to the passing gravitational wave,
because they also move in the gravitational field of the solar system bodies,
as in the case of the LISA spacecraft,
or are fixed to the surface of the Earth,
as in the case of Earth-based laser interferometers or resonant bar detectors.
Let us choose the origin O of the TT coordinate system
employed in Section~\ref{subsection:Doppler_shift}
to coincide with the \emph{solar system barycenter} (SSB).
The motion of the detector with respect to the SSB
will modulate the  gravitational-wave signal registered by the detector.
One can show that as far as the velocities of the particles
(modeling the detector's parts) with respect to the SSB are \emph{non-relativistic},
which is the case for all existing or planned detectors,
Eqs.~\eqref{r2.11} can still be used,
provided the 3-vectors ${\mathbf{x}}_a$ and ${\mathbf{n}}_a$ ($a=1,2,3$)
will be interpreted as made of the time-dependent components
describing the motion of the particles with respect to the SSB.

It is often convenient to introduce the \emph{proper reference frame}
of the detector with coordinates $(\hat{x}^\alpha)$.
Because the motion of this frame with respect to the SSB is non-relativistic,
we can assume that the transformation between the SSB-related coordinates $(x^\alpha)$
and the detector's proper reference frame coordinates $(\hat{x}^\alpha)$ has the form
\begin{equation}
\label{SSBtoD}
\hat{t} = t, \qquad
\hat{x}^i(t,x^k) = \hat{x}^i_{\widehat{O}}(t) + \mathsf{O}^i_j(t)\,x^j,
\end{equation}
where the functions $\hat{x}^i_{\widehat{O}}(t)$ describe
the motion of the origin ${\widehat{O}}$ of the proper reference frame
with respect to the SSB, and the functions $\mathsf{O}^i_j(t)$
account for the different orientations of the spatial axes
of the two reference frames. One can compute some of the quantities
entering Eqs.~\eqref{r2.11} in the detector's coordinates
rather than in the TT coordinates. For instance, the matrix
$\widehat{\mathsf{H}}$ of the wave-induced spatial metric perturbation in
the detector's coordinates is related to the matrix $\mathsf{H}$ of the
spatial metric perturbation produced by the wave in the TT coordinate
system through equation
\begin{equation}
\label{hatH}
\widehat{\mathsf{H}}(t) = (\mathsf{O}(t)^{-1})^{\mathsf{T}} \cdot \mathsf{H}(t) \cdot
\mathsf{O}(t)^{-1},
\end{equation}
where the matrix $\mathsf{O}$ has elements $\mathsf{O}^i_j$.
If the transformation matrix $\mathsf{O}$ is \emph{orthogonal},
then $\mathsf{O}^{-1}=\mathsf{O}^{\mathsf{T}}$, and Eq.~(\ref{hatH}) simplifies to
\begin{equation}
\label{HhatH}
\widehat{\mathsf{H}}(t) = {\mathsf{O}}(t)\cdot{\mathsf{H}}(t)\cdot{\mathsf{O}}(t)^{\mathsf{T}}.
\end{equation}
See~\cite{bonazzola-96,giampieri-97,jaranowski-98,krolak-04}
for more details.

For a \emph{ground-based laser-interferometric detector}, the long-wavelength
approximation can be employed (however, see~\cite{BaskaranGrishchuk2004,RakhmanovRomanoWhelan2008,Rakhmanov2009}
for a discussion of importance of high-frequency corrections,
which modify the interferometer response function
computed within the long-wavelength approximation).
In the case of an interferometer
in a standard Michelson and equal-arm configuration
(such configurations can be represented by Figure~\ref{fig:3particles}
with the particle 1 corresponding to the corner station of the interferometer
and with $L_2=L_3=L$), the observed relative frequency shift
$\Delta\nu(t)/\nu_0$ is equal to the difference of the round-trip frequency shifts
in the two detector's arms~\cite{tinto-99}:
\begin{equation}
\label{eq:dfM}
\frac{\Delta\nu(t)}{\nu_0} = y_{131}(t) - y_{121}(t).
\end{equation}
Let $(x_\mathrm{d},y_\mathrm{d},z_\mathrm{d})$ be the components
(with respect to the TT coordinate system)
of the 3-vector $\mathbf{r}_\mathrm{d}$ connecting the origin of the TT coordinate system
with the corner station of the interferometer.
Then $\mathsf{x}_1=(x_\mathrm{d},y_\mathrm{d,}z_\mathrm{d})^\mathsf{T}$,
$\mathsf{k}^\mathsf{T}\cdot\mathsf{x}_1=-z_\mathrm{d}$, and,
making use of Eq.~\eqref{r1.lw3},
the relative frequency shift~\eqref{eq:dfM} can be written as
\begin{equation}
\label{lwa}
\frac{\Delta\nu(t)}{\nu_0} = \frac{L}{c} \left( \mathsf{n}_2^\mathsf{T}
\cdot \dot{\mathsf{H}}\Big(t-\frac{z_\mathrm{d}}{c}\Big) \cdot \mathsf{n}_2
- \mathsf{n}_3^\mathsf{T}
\cdot \dot{\mathsf{H}}\Big(t-\frac{z_\text{d}}{c}\Big) \cdot \mathsf{n}_3 \right).
\end{equation}
The difference $\Delta\phi(t)$ of the phase fluctuations measured,
say, by a photo detector, is related to the corresponding relative
frequency fluctuations $\Delta\nu(t)$ by
\begin{equation}
\label{PhaseChange1}
\frac{\Delta\nu(t)}{\nu_0} = \frac{1}{2\pi\nu_0}
\frac{{\mathrm{d}}\Delta\phi(t)}{{\mathrm{d}}t}.
\end{equation}
One can integrate Eq.~\eqref{lwa}
to write the phase change $\Delta\phi(t)$ as
\begin{equation}
\label{PhaseChange2}
\Delta\phi(t) = 4 \pi \nu_0 L \, h(t),
\end{equation}
where the dimensionless function $h$,
\begin{equation}
\label{IFresponse1}
h(t) := \frac{1}{2} \left( \mathsf{n}_2^\mathsf{T}
\cdot \mathsf{H}\Big(t-\frac{z_\mathrm{d}}{c}\Big) \cdot \mathsf{n}_2
- \mathsf{n}_3^\mathsf{T}
\cdot \mathsf{H}\Big(t-\frac{z_\text{d}}{c}\Big) \cdot \mathsf{n}_3
\right),
\end{equation}
is the \emph{response function} of the interferometric detector
to a plane gravitational wave
in the long-wavelength approximation.
To get Eqs.~\eqref{PhaseChange2}\,--\,\eqref{IFresponse1}
directly from Eqs.~\eqref{lwa}\,--\,\eqref{PhaseChange1}
one should assume that the quantities $\mathsf{n}_2$, $\mathsf{n}_3$, and $z_\text{d}$
[entering Eq.~\eqref{lwa}] do not depend on time $t$.
But the formulae~\eqref{PhaseChange2}\,--\,\eqref{IFresponse1}
can also be used in the case when those quantities are time dependent,
provided the velocities $\dot{\mathbf{x}}_a$ of the detector's parts
with respect to the SSB are non-relativistic.
The error we make in such cases is on the order of $\mathcal{O}(hv)$,
where $v$ is a typical value of the velocities $\dot{\mathbf{x}}_a$.
Thus, the response function of the Earth-based interferometric detector equals
\begin{equation}
\label{IFresponse2}
h(t) = \frac{1}{2} \left( \mathsf{n}_2(t)^\mathsf{T}
\cdot \mathsf{H}\Big(t-\frac{z_\mathrm{d}(t)}{c}\Big) \cdot \mathsf{n}_2(t)
- \mathsf{n}_3(t)^\mathsf{T}
\cdot \mathsf{H}\Big(t-\frac{z_\text{d}(t)}{c}\Big) \cdot \mathsf{n}_3(t)
\right),
\end{equation}
where all quantities here are computed in the SSB-related TT coordinate system.
The same response function can be written in terms of a detector's
proper-reference-frame quantities as follows
\begin{equation}
\label{IFresponse3}
h(t) = \frac{1}{2} \left( \hat{\mathsf{n}}_2^\mathsf{T}
\cdot \hat{\mathsf{H}}\Big(t-\frac{z_\mathrm{d}(t)}{c}\Big) \cdot \hat{\mathsf{n}}_2
- \hat{\mathsf{n}}_3^\mathsf{T}
\cdot \hat{\mathsf{H}}\Big(t-\frac{z_\text{d}(t)}{c}\Big) \cdot \hat{\mathsf{n}}_3
\right),
\end{equation}
where the matrices $\hat{\mathsf{H}}$ and $\mathsf{H}$ are related to each other
by means of formula~\eqref{HhatH}. In Eq.~\eqref{IFresponse3}
the proper-reference-frame components $\hat{\mathsf{n}}_2$ and $\hat{\mathsf{n}}_3$
of the unit vectors directed along the interferometer arms
can be treated as constant (i.e., time independent) quantities.

From Eqs.~\eqref{IFresponse2} and \eqref{r2.2} it follows
that the response function $h$ is a linear combination
of the two wave polarizations $h_+$ and $h_\times$,
so it can be written as
\be
\label{IFresponse4}
h(t) = F_+(t)\,h_+\Big(t-\frac{z_\mathrm{d}(t)}{c}\Big)
+ F_\times(t)\,h_\times\Big(t-\frac{z_\mathrm{d}(t)}{c}\Big).
\ee
The functions $F_+$ and $F_\times$ are the interferometric \emph{beam-pattern functions}.
They depend on the location of the detector on Earth,
the position of the gravitational-wave source in the sky,
and the polarization angle of the wave
(this angle describes the orientation,
with respect to the detector, of the axes relative to which
the plus and cross polarizations of the wave are defined,
see, e.g., Figure~9.2 in~\cite{thorne-87}).
Derivation of the explicit formulae
for the interferometric beam patterns $F_+$ and $F_\times$
can be found, e.g., in Appendix~C of~\cite{JaranowskiKrolak2009}.

In the long-wavelength approximation,
the response function of the interferometric detector
can be derived directly from the equation of geodesic deviation~\cite{SchutzTinto1987}.
Then the response is defined as the relative difference
between the wave-induced changes of the proper lengths of the two arms,
i.e., $h(t):=(\Delta\hat{L}_2(t)-\Delta\hat{L}_3(t))/\hat{L}_0$,
where $\hat{L}_0+\Delta\hat{L}_2(t)$ and $\hat{L}_0+\Delta\hat{L}_3(t)$
are the instantaneous values of the proper lengths of the interferometer's arms
and $\hat{L}_0$ is the unperturbed proper length of these arms.

In the case of an Earth-based resonant-bar detector
the long-wavelength approximation is very accurate
and the dimensionless detector's response function
can be defined as $h(t):=\Delta\hat{L}(t)/\hat{L}_0$,
where $\Delta\hat{L}(t)$ is the wave-induced change
of the proper length $\hat{L}_0$ of the bar.
The response function $h$ can be computed
in terms of the detector's proper-reference-frame quantities
from the formula (see, e.g., Section~9.5.2 in~\cite{thorne-87})
\begin{equation}
\label{BarResponse}
h(t) = \hat{\mathsf{n}}^{\mathsf{T}}
\cdot \hat{\mathsf{H}}\Big(t-\frac{z_\mathrm{d}(t)}{c}\Big)
\cdot \hat{\mathsf{n}},
\end{equation}
where the column matrix $\hat{\mathsf{n}}$ consists of the components
(computed in the proper reference frame of the detector)
of the unit vector ${\mathbf{n}}$
directed along the symmetry axis of the bar.
The response function~\eqref{BarResponse} can be written
as a linear combination of the wave polarizations $h_+$ and $h_\times$,
i.e., the formula~\eqref{IFresponse4} is also valid for the resonant-bar response
function but with some \emph{bar-pattern functions} $F_+$ and $F_\times$
different from the interferometric beam-pattern functions. Derivation
of the explicit form of the bar patterns can be found, e.g.,
in Appendix~C of~\cite{JaranowskiKrolak2009}.

\subsection{General form of the response function}

In many cases of interest the response function of the detector to a plane gravitational wave
can be written as a linear combination of $n$ waveforms $h_k(t;\bxi)$
[which all depend on the same set of parameters $\bxi=(\xi_1,\ldots,\xi_m)$]
with constant amplitudes $a_k$ ($k=1,\ldots,n$),
\begin{equation}
\label{eq:gsig}
h(t;\btheta) = \sum^n_{k=1} a_k\,h_k(t;\bxi),
\end{equation}
where the vector $\btheta$ collects all the signal's parameters.
It is convenient to introduce column matrices
\be
\mathsf{a} := \begin{pmatrix}a_1 \\ \vdots \\ a_n \end{pmatrix},
\quad
\mathsf{h}(t;\bxi) := \begin{pmatrix}h_1(t;\bxi) \\ \vdots \\ h_n(t;\bxi) \end{pmatrix}.
\ee
Then one can briefly write $\btheta=(\mathsf{a},\bxi)$
and the response~\eqref{eq:gsig} can be written as
\begin{equation}
\label{gsig2}
h(t;\btheta)
= \mathsf{a}^\mathsf{T}\cdot\mathsf{h}(t;\bxi).
\end{equation}
The functions $h_k$ ($k=1,\ldots,n$) are independent of the parameters $\mathsf{a}$.
The parameters $\mathsf{a}$ are called \emph{extrinsic} (or \emph{amplitude}) parameters
whereas the parameters $\bxi$ are called \emph{intrinsic}.

Eq.~(\ref{gsig2}) with $n=4$ is a model of the response
of the space-based detector LISA to gravitational waves from a binary
system~\cite{krolak-04}. Also for $n=4$ the same equation models
the response of a ground-based detector to a continuous source of
gravitational waves like a rotating neutron
star~\cite{jaranowski-98}. For ground-based detectors
the long-wavelength approximation can be applied
and within this approximation the functions $h_k$ ($k=1,\ldots,4$) are given by
\begin{equation}
  \label{eq:LWsig}
  \begin{array}{rcl}
    h_1(t;\bxi) &=& u(t;\bxi) \cos\phi(t;\bxi), \\[0.5em]
    h_2(t;\bxi) &=& v(t;\bxi) \cos\phi(t;\bxi), \\[0.5em]
    h_3(t;\bxi) &=& u(t;\bxi) \sin\phi(t;\bxi), \\[0.5em]
    h_4(t;\bxi) &=& v(t;\bxi) \sin\phi(t;\bxi),
  \end{array}
\end{equation}
where $\phi(t;\bxi)$ is the phase modulation of the signal and
$u(t;\bxi)$ and $v(t;\bxi)$ are slowly varying amplitude
modulations. The gravitational-wave signal from spinning
neutron stars may consist of several components of the
form~(\ref{gsig2}).
For short observation times over which
the amplitude modulation functions are nearly constant,
the response of the ground-based detector
can further be approximated by
\begin{equation}
  \label{eq:chsig}
  h(t; A_0,\phi_0,\bxi) =
  A_0 \, g(t;\bxi) \cos\left(\phi(t;\bxi)-\phi_0\right),
\end{equation}
where $A_0$ and $\phi_0$ are constant amplitude and initial phase,
respectively, and $g(t;\bxi)$ is a slowly varying function of time.
Eq.~(\ref{eq:chsig}) is a good model for the response of
a detector to the gravitational wave from a coalescing compact
binary system~\cite{thorne-87, blanchet-06}. We would like to stress
that not all deterministic gravitational-wave signals may be cast into
the general form~(\ref{gsig2}).

\newpage

\section{Statistical Theory of Signal Detection}
\label{sec:statistical-theory}

The gravitational-wave signal will be buried in the noise of the
detector and the data from the detector will be a random (or
stochastic) process. Consequently, the problem of extracting the signal
from the noise is a statistical one. The basic idea behind signal
detection is that the presence of the signal changes the statistical
characteristics of the data $x$, in particular its probability
distribution. When the signal is absent the data have probability
density function (pdf) $p_0(x)$, and when the signal is present the
pdf is $p_1(x)$.

A thorough introduction to probability theory and mathematical statistics
can be found, e.g., in~\cite{Fisz1967, KATSi1994, KATSiiA1999, KATSiiB2004}.
A full exposition of statistical theory of signal detection that is
only outlined here can be found in the monographs~\cite{woodward-53,
  kotelnikov-59, wainstein-62, trees-68, donough-95, helstrom-95,
  poor-95}. A general introduction to stochastic processes is given
in~\cite{wong-83} and advanced treatment of the subject can be found
in~\cite{liptser-77, wong-85}. A concise introduction to the statistical
theory of signal detection and time series analysis is contained in
Chapters~3 and 4 of~\cite{JaranowskiKrolak2009}.

\subsection{Hypothesis testing}

The problem of detecting the signal in noise can be posed as a
statistical hypothesis testing problem. The \emph{null hypothesis}
$H_0$ is that the signal is absent from the data and the
\emph{alternative hypothesis} $H_1$ is that the signal is present. A
\emph{hypothesis test} (or \emph{decision rule}) $\delta$ is a
partition of the observation set into two subsets, ${\cal R}$ and its
complement ${\cal R}'$. If data are in ${\cal R}$ we accept the null
hypothesis, otherwise we reject it. There are two kinds of errors
that we can make. A type~I error is choosing hypothesis $H_1$ when
$H_0$ is true and a type~II error is choosing $H_0$ when $H_1$ is
true. In signal detection theory the probability of a type I error is
called the \emph{false alarm probability}, whereas the probability of
a type II error is called the \emph{false dismissal probability}.
$1-\mbox{(false dismissal probability)}$
is the \emph{probability of detection} of the signal.
In hypothesis testing theory, the probability
of a type I error is called the \emph{significance of the test},
whereas $1-\mbox{(probability of type II error)}$ is called the
\emph{power of the test}.

The problem is to find a test that is in some way optimal.
There are several approaches to finding such a test.
The subject is covered in detail in many books on statistics,
for example, see~\cite{kendall-95,Fisz1967,Lehmann1959,LehmannRomano2005}.


\subsubsection{Bayesian approach}

In the Bayesian approach we assign costs to our decisions; in
particular we introduce positive numbers $C_{ij}$, $i, j = 0, 1$,
where $C_{ij}$ is the cost incurred by choosing hypothesis $H_i$ when
hypothesis $H_j$ is true. We define the \emph{conditional risk} $R$ of
a decision rule $\delta$ for each hypothesis as
\begin{equation}
  R_j(\delta) = C_{0j}P_j({\cal R}) + C_{1j}P_j({\cal R}'),
  \qquad
  j = 0,1,
\end{equation}
where $P_j$ is the probability distribution of the data when
hypothesis $H_j$ is true. Next, we assign probabilities $\pi_0$ and
$\pi_1=1-\pi_0$ to the occurrences of hypotheses $H_0$ and $H_1$,
respectively. These probabilities are called \emph{a priori
  probabilities} or \emph{priors}. We define the \emph{Bayes risk} as
the overall average cost incurred by the decision rule $\delta$:
\begin{equation}
  r(\delta) = \pi_0 R_0(\delta) + \pi_1 R_1(\delta).
\end{equation}
Finally we define the \emph{Bayes rule} as the rule that minimizes the
Bayes risk $r(\delta)$.


\subsubsection{Minimax approach}

Very often in practice we do not have control over or access
to the mechanism generating the state of nature and we are not
able to assign priors to various hypotheses. In such a case one
criterion is to seek a decision rule that minimizes, over all
$\delta$, the maximum of the conditional risks, $R_0(\delta)$ and
$R_1(\delta)$. A decision rule that fulfills that criterion is
called a \emph{minimax rule}.


\subsubsection{Neyman--Pearson approach}

In many problems of practical interest the imposition of a specific
cost structure on the decisions made is not possible or desirable. The
Neyman--Pearson approach involves a trade-off between the two types of
errors that one can make in choosing a particular hypothesis. The
Neyman--Pearson design criterion is to maximize the power of the test
(probability of detection) subject to a chosen significance of the
test (false alarm probability).


\subsubsection{Likelihood ratio test}

It is remarkable that all three very different approaches -- Bayesian,
minimax, and Neyman--Pearson -- lead to the same test called the
\emph{likelihood ratio test}~\cite{davis-89}. The likelihood ratio
$\Lambda$ is the ratio of the pdf when the signal is present to the
pdf when it is absent:
\begin{equation}
  \label{eq:LR}
  \Lambda(x) := \frac{p_1(x)}{p_0(x)}.
\end{equation}
We accept the hypothesis $H_1$ if $\Lambda>k$, where $k$ is the
threshold that is calculated from the costs $C_{ij}$, priors $\pi_i$,
or the significance of the test depending on which approach is being used.


\subsection{The matched filter in Gaussian noise}

Let $h$ be the gravitational-wave signal we are looking for
and let $n$ be the detector's noise.
For convenience we assume that the signal $h$
is a continuous function of time $t$
and that the noise $n$ is a continuous random process.
Results for the discrete-in-time data that we have in practice
can then be obtained by a suitable sampling
of the continuous-in-time expressions.
Assuming that the noise is \emph{additive}
the data $x$ can be written as
\begin{equation}
\label{eq:D}
x(t) = n(t) + h(t).
\end{equation}
The \emph{autocorrelation function}
of the noise $n$ is defined as
\begin{equation}
\label{Kn}
K_n(t,t') := \mE\left[n(t)n(t')\right],
\end{equation}
where E denotes the expectation value.

Let us further assume that the detector's noise $n$ is a \emph{zero-mean}
and \emph{Gaussian} random process.
It can then be shown that the logarithm of the likelihood function
is given by the following \emph{Cameron--Martin} formula
\begin{equation}
\label{CMformula}
\log\Lambda[x] = \int_0^\To q(t)\,x(t)\,\md t
- \frac{1}{2} \int_0^\To q(t)\,h(t)\,\md t,
\end{equation}
where $\langle0;\To\rangle$ is the time interval over which the data was collected
and the function $q$ is the solution of the integral equation
\begin{equation}
\label{IntEq}
h(t) = \int_0^\To K_n(t,t')q(t')\,\md t.
\end{equation}

For \emph{stationary} noise, its autocorrelation function~\eqref{Kn}
depends on times $t$ and $t'$ only through the difference $t-t'$.
It implies that there exists then an \emph{even} function $\kappa_n$ of one variable such that
\begin{equation}
\mE[n(t)n(t')] = \kappa_n(t-t').
\end{equation}
Spectral properties of stationary noise are described by its
\emph{one-sided spectral density},
defined for non-negative frequencies $f\ge0$ through relation
\begin{equation}
S_n(f) = 2 \int_{-\infty}^\infty \kappa_n(t) e^{2\pi i f t} \md t.
\end{equation}
For negative frequencies $f<0$, by definition, $S_n(f)=0$. The spectral density $S_n$
can also be determined by correlations between the Fourier components
of the detector's noise
\begin{equation}
\label{eq:SD}
\mE\left[\tilde{n}(f)\tilde{n}^{*}(f')\right]
= \frac{1}{2} S_n(|f|) \delta(f-f'), \quad
-\infty<f,f'<\infty.
\end{equation}

For the case of stationary noise with one-sided spectral density $S_n$,
it is convenient to define the scalar product $(x|y)$
between any two waveforms $x$ and $y$,
\begin{equation}
\label{eq:SP}
(x|y) := 4 \Re \int^\infty_0
\frac{\tilde{x}(f)\tilde{y}^{*}(f)}{S_n(f)}\,\md f,
\end{equation}
where $\Re$ denotes the real part of a complex expression,
the tilde denotes the Fourier transform,
and the asterisk is complex conjugation.
Making use of this scalar product,
the log likelihood function~\eqref{CMformula} can be written as
\begin{equation}
  \label{eq:L}
  \log\Lambda[x] = (x|h) - \frac{1}{2}(h|h).
\end{equation}
From the expression~\eqref{eq:L} we see immediately
that the likelihood ratio test consists of correlating the data $x$
with the signal $h$ that is present in the noise
and comparing the correlation to a threshold.
Such a correlation is called the \emph{matched filter}.
The matched filter is a \emph{linear} operation on the data.

An important quantity is the optimal \emph{signal-to-noise ratio}
$\rho$ defined by
\begin{equation}
\label{eq:d}
\rho := \sqrt{(h|h)}.
\end{equation}
By means of Eq.~\eqref{eq:SP} it can be written as
\begin{equation}
\label{eq:d2}
\rho^2 = 4 \int^\infty_0 \frac{|\tilde{h}(f)|^2}{S_n(f)}\,\md f.
\end{equation}
We see in the following that $\rho$ determines the probability
of detection of the signal. The higher the signal-to-noise ratio
the higher the probability of detection.

An interesting property of the matched filter is that it maximizes the
signal-to-noise ratio over all linear filters~\cite{davis-89}. This
property is independent of the probability distribution of the noise.


\subsection{Parameter estimation}

Very often we know the waveform of the signal that we are
searching for in the data in terms of a finite number of unknown
parameters. We would like to find optimal procedures of
estimating these parameters. An estimator of a parameter $\theta$
is a function $\hat{\theta}(x)$ that assigns to each data set the
``best'' guess of the true value of $\theta$. Note that because
$\hat{\theta}(x)$ depends on the random data it is a random
variable. Ideally we would like our estimator to be (i)
\emph{unbiased}, i.e., its expectation value to be equal to the true
value of the parameter, and (ii) \emph{of minimum variance}. Such
estimators are rare and in general difficult to find. As in the
signal detection there are several approaches to the parameter
estimation problem.
The subject is exposed in detail in~\cite{Lehmann1983,LehmannCasella1998}.
See also~\cite{zielinski-97} for a concise account.


\subsubsection{Bayesian estimation}

We assign a cost function $C(\theta',\theta)$ of estimating the true
value of $\theta$ as $\theta'$. We then associate with an estimator
$\hat{\theta}$ a conditional risk or cost averaged over all
realizations of data $x$ for each value of the parameter $\theta$:
\begin{equation}
R_{\theta}(\hat{\theta})
= \mathrm{E}_{\theta}[C(\hat{\theta},\theta)]
= \int_X C\big(\hat{\theta}(x),\theta\big)\,p(x,\theta)\,\md x,
\end{equation}
where $X$ is the set of observations and $p(x,\theta)$ is the joint
probability distribution of data $x$ and parameter $\theta$. We
further assume that there is a certain \emph{a priori} probability
distribution $\pi(\theta)$ of the parameter $\theta$. We then define
the \emph{Bayes estimator} as the estimator that minimizes the average
risk defined as
\begin{equation}
\label{eq:avR}
r(\hat{\theta}) = \mathrm{E}[R_{\theta}(\hat{\theta})]
= \int_X  \int_\Theta C \big(\hat{\theta}(x),\theta\big)\,
p(x,\theta)\, \pi(\theta) \,\md\theta\,\md x,
\end{equation}
where E is the expectation value with respect to an \emph{a priori}
distribution $\pi$, and $\Theta$ is the set of observations of the
parameter $\theta$. It is not difficult to show that for a commonly
used cost function
\begin{equation}
  C(\theta',\theta) = (\theta'-\theta)^2,
\end{equation}
the Bayesian estimator is the conditional mean of the parameter
$\theta$ given data $x$, i.e.,
\begin{equation}
  \label{eq:correct}
  \hat{\theta}(x) = \mathrm{E}[\theta|x]
  = \int_{\Theta} \theta\,p(\theta|x)\,\md\theta,
\end{equation}
%
where $p(\theta|x)$ is the conditional probability density of
parameter $\theta$ given the data $x$.


\subsubsection{Maximum \emph{a posteriori} probability estimation}

Suppose that in a given estimation problem we are not able to assign a
particular cost function  $C(\theta',\theta)$. Then a natural choice
is a uniform cost function equal to 0 over a certain interval
$I_{\theta}$ of the parameter $\theta$. From Bayes
theorem~\cite{bayes-1763} we have
\begin{equation}
  p(\theta|x) = \frac{p(x,\theta) \pi(\theta)}{p(x)},
\end{equation}
where $p(x)$ is the probability distribution of data $x$. Then, {from}
Eq.~(\ref{eq:avR}) one can deduce that for each data point $x$ the
Bayes estimate is any value of $\theta$ that maximizes the conditional
probability $p(\theta|x)$. The density $p(\theta|x)$ is also called
the \emph{a posteriori} probability density of parameter $\theta$ and
the estimator that maximizes $p(\theta|x)$ is called the \emph{maximum
  a posteriori} (MAP) estimator. It is denoted by
$\hat{\theta}_\mathrm{MAP}$. We find that the MAP estimators are
solutions of the following equation
\begin{equation}
  \frac{\partial \log p(x,\theta)}{\partial \theta} =
  - \frac{\partial \log \pi(\theta)}{\partial \theta},
\end{equation}
which is called the \emph{MAP equation}.


\subsubsection{Maximum likelihood estimation}

Often we do not know the \emph{a priori} probability density of a given
parameter and we simply assign to it a uniform probability. In such a
case maximization of the \emph{a posteriori} probability is equivalent to
maximization of the probability density $p(x,\theta)$ treated as a
function of $\theta$. We call the function $l(\theta, x) :=
p(x,\theta)$ the likelihood function and the value of the parameter
$\theta$ that maximizes $l(\theta, x)$ the \emph{maximum likelihood}
(ML) estimator. Instead of the function $l$ we can use the function
$\Lambda(\theta, x) = l(\theta, x)/p(x)$ (assuming that $p(x) >
0$). $\Lambda$ is then equivalent to the likelihood ratio [see
Eq.~(\ref{eq:LR})] when the parameters of the signal are
known. Then the ML estimators are obtained by solving the equation
\begin{equation}
\frac{\partial\log\Lambda(\theta, x)}{\partial\theta} = 0,
\end{equation}
which is called the \emph{ML equation}.


\subsection{Fisher information and Cram\`er--Rao bound}

It is important to know how good our estimators are. We would like our
estimator to have as small a variance as possible. There is a useful
lower bound on variances of the parameter estimators called the \emph{Cram\`er--Rao bound},
which is expressed in terms of the \emph{Fisher information matrix} $\Gamma(\btheta)$.
For the signal $h(t;\btheta)$, which depends on $K$ parameters
collected into the vector $\btheta=(\theta_1,\ldots,\theta_K)$,
the components of the matrix $\Gamma(\btheta)$ are defined as
\begin{equation}
  \label{eq:GA}
  \Gamma(\btheta)_{ij} :=
  \mathrm{E} \left[ \frac{\partial\log\Lambda[x;\btheta]}{\partial\theta_i}
  \frac{\partial\log\Lambda[x;\btheta]}{\partial\theta_j} \right] =
  -\mathrm{E} \left[ \frac{\partial^2\log\Lambda[x;\btheta]}
  {\partial\theta_i\,\partial\theta_j} \right],
\quad i,j=1,\ldots,K.
\end{equation}
The Cram\`er--Rao bound states that for \emph{unbiased} estimators
the \emph{covariance matrix} $\mathsf{C}(\btheta)$ of the estimators $\btheta$
fulfills the inequality
\be
\mathsf{C}(\btheta) \geq \Gamma(\btheta)^{-1}.
\ee
(The inequality $\mathsf{A}\geq\mathsf{B}$ for matrices means
that the matrix $\mathsf{A}-\mathsf{B}$ is nonnegative definite.)

A very important property of the ML estimators is that asymptotically
(i.e., for a signal-to-noise ratio tending to infinity) they are (i)
unbiased, and (ii) they have a Gaussian distribution with covariance
matrix equal to the inverse of the Fisher information matrix.

In the case of Gaussian noise the components of the Fisher matrix
are given by
\begin{equation}
\label{eq:GAg}
\Gamma(\btheta)_{ij} = \left( \frac{\partial h(t;\btheta)}{\partial\theta_i}
\bigg| \frac{\partial h(t;\btheta)}{\partial\theta_j} \right),
\quad i,j=1,\ldots,K,
\end{equation}
where the scalar product $(\cdot|\cdot)$ is defined in Eq.~\eqref{eq:SP}.

\newpage

\section{Maximum-likelihood Detection in Gaussian Noise}
\label{sec:Fisher-matrix-Gaussian-case}

In this section, we study the detection
of a deterministic gravitational-wave signal $h(t;\btheta)$
of the general form given by Eq.~\eqref{gsig2}
and the estimation of its parameters $\btheta$ using the maximum-likelihood (ML) principle.
We assume that the noise $n(t)$ in the detector
is a \emph{zero-mean}, \emph{Gaussian}, and \emph{stationary} random process.
The data $x$ in the detector,
in the case when the gravitational-wave signal $h(t;\btheta)$ is present,
is $x(t;\btheta) = n(t) + h(t;\btheta)$.
The parameters $\btheta=(\mathsf{a},\bxi)$ of the signal~\eqref{gsig2}
split into extrinsic (or amplitude) parameters~$\mathsf{a}$ and intrinsic ones~$\bxi$.


\subsection{The $\F$-statistic}

For the gravitational-wave signal $h(t;\mathsf{a},\bxi)$
of the form given in Eq.~(\ref{gsig2})
the log likelihood function~\eqref{eq:L} can be written as
\begin{equation}
\label{LF1}
\log\Lambda[x;\mathsf{a},\bxi] = {\mathsf a}^\mathsf{T}\cdot{\mathsf N}[x;\bxi]
- \frac{1}{2}\,{\mathsf a}^\mathsf{T}\cdot\mathsf{M}(\bxi)\cdot{\mathsf a},
\end{equation}
where the components of the column $n\times1$ matrix ${\mathsf N}$
and the square $n\times n$ matrix $\mathsf{M}$ are given by
\begin{equation}
\label{eq:M}
N_k[x;\bxi] := (x|h_k(t;\bxi)), \quad
M_{kl}(\bxi) := (h_k(t;\bxi)|h_l(t;\bxi)), \qquad k,l=1,\ldots,n.
\end{equation}
The ML equations for the extrinsic parameters ${\mathsf a}$,
$\partial\log\Lambda[x;\mathsf{a},\bxi]/\partial\mathsf{a}=\boldsymbol{0}$,
can be solved explicitly to show that the ML estimators $\mathsf{\hat{a}}$
of the parameters $\mathsf{a}$ are given by
\begin{equation}
\label{eq:aMLE}
\hat{\mathsf{a}}[x;\bxi] = \mathsf{M}(\bxi)^{-1}\cdot{\mathsf N}[x;\bxi].
\end{equation}
Replacing the extrinsic parameters $\mathsf{a}$ in Eq.~\eqref{LF1}
by their ML estimators $\hat{\mathsf{a}}$,
we obtain the reduced log likelihood function,
\begin{equation}
\label{eq:FS}
\F[x;\bxi] := \log\Lambda\big[x;\hat{\mathsf{a}}[x;\bxi],\bxi\big]
= \frac{1}{2}\,{\mathsf N}[x;\bxi]^\mathsf{T}
\cdot\mathsf{M}(\bxi)^{-1}\cdot{\mathsf N}[x;\bxi],
\end{equation}
that we call the \emph{$\F$-statistic}. The $\F$-statistic depends
nonlinearly on the intrinsic parameters $\bxi$ of the signal,
it does not depend on the extrinsic parameters $\mathsf{a}$.

The procedure to detect the gravitational-wave signal of the form~\eqref{gsig2}
and estimate its parameters consists of two parts.
The first part is to find the (local) maxima of
the $\F$-statistic~\eqref{eq:FS} in the intrinsic parameters space.
The ML estimators $\hat{\bxi}$ of the intrinsic parameters $\bxi$
are those values of $\bxi$ for which the $\F$-statistic attains a maximum.
The second part is to calculate the estimators $\hat{\mathsf{a}}$
of the extrinsic parameters $\mathsf{a}$ from the analytic formula~(\ref{eq:aMLE}),
where the matrix $\mathsf{M}$ and the correlations $\mathsf{N}$
are calculated for the parameters $\bxi$ replaced by their ML estimators $\hat{\bxi}$
obtained from the first part of the analysis. We call this procedure
the \emph{maximum likelihood detection}. See Section~\ref{sec:alg} for
a discussion of the algorithms to find the (local) maxima of the
$\F$-statistic.

\subsubsection{Targeted searches}
\label{sec:targeted-searches}

The $\F$-statistic can also be used in the case when the intrinsic parameters are known.
An example of such an analysis called a \emph{targeted search}
is the search for a gravitational-wave signal from a known pulsar.
In this case assuming that gravitational-wave emission follows the radio timing,
the phase of the signal is known from pulsar observations and the only unknown parameters
of the signal are the amplitude (or extrinsic) parameters $\mathsf{a}$ [see Eq.~\eqref{eq:gsig}].
To detect the signal one calculates the $\F$-statistic
for the known values of the intrinsic parameters and compares it to a threshold~\cite{JK2010}.
When a statistically-significant signal is detected,
one then estimates the amplitude parameters from the analytic formulae~\eqref{eq:aMLE}.

In~\cite{PrixKrishnan2009} it was shown that the maximum-likelihood $\F$-statistic
can be interpreted as a Bayes factor with a simple, but unphysical, amplitude prior
(and an additional unphysical sky-position weighting).
Using a more physical prior based on an isotropic probability
distribution for the unknown spin-axis orientation of emitting systems,
a new detection statistic (called the $\B$-statistic) was obtained.
Monte Carlo simulations for signals with random (isotropic) spin-axis orientations show
that the $\B$-statistic is more powerful
(in terms of its expected detection probability) than the $\F$-statistic.
A modified version of the $\F$-statistic that can be more powerful than the original one
has been studied in~\cite{AstoneOthers2010}.

\subsection{Signal-to-noise ratio and the Fisher matrix}

The detectability of the signal $h(t;\btheta)$
is determined by the signal-to-noise ratio $\rho$.
In general it depends on all the signal's parameters $\btheta$
and can be computed from [see Eq.~\eqref{eq:d}]
\be
\label{snr1}
\rho(\btheta) = \sqrt{(h(t;\btheta)|h(t;\btheta))}.
\ee
The signal-to-noise ratio for the signal~\eqref{gsig2}
can be written as
\be
\label{snr2}
\rho(\mathsf{a},\bxi) = \sqrt{\mathsf{a}^\mathsf{T}
\cdot \mathsf{M}(\bxi) \cdot\mathsf{a}},
\ee
where the components of the matrix $\mathsf{M}(\bxi)$
are defined in Eq.~\eqref{eq:M}.

The accuracy of estimation of the signal's parameters
is determined by Fisher information matrix $\Gamma$.
The components of $\Gamma$ in the case of the Gaussian noise
can be computed from Eq.~\eqref{eq:GAg}.
For the signal given in Eq.~\eqref{gsig2} the signal's parameters
(collected into the vector $\btheta$) split into extrinsic and intrinsic
parameters: $\btheta=(\mathsf{a},\bxi)$, where
$\mathsf{a}=(a_1,\ldots,a_n)$ and $\bxi=(\xi_1,\ldots,\xi_m)$.
It is convenient to distinguish between extrinsic and intrinsic parameter indices.
Therefore, we use calligraphic lettering to denote the intrinsic parameter indices:
$\xi_\mathcal{A}$, $\mathcal{A}=1,\ldots,m$.
The matrix $\Gamma$ has dimension $(n+m)\times(n+m)$
and it can be written in terms of four block matrices
for the two sets of the parameters $\mathsf{a}$ and $\bxi$,
\be
\label{G}
\Gamma(\mathsf{a},\bxi) = \begin{pmatrix}
\Gamma_{\mathsf{a}\mathsf{a}}(\bxi)
& \Gamma_{\mathsf{a}\bxi}(\mathsf{a},\bxi)
\\\noalign{\smallskip}
\Gamma_{\mathsf{a}\bxi}(\mathsf{a},\bxi)^\mathsf{T}
& \Gamma_{\bxi\bxi}(\mathsf{a},\bxi)
\end{pmatrix},
\ee
where $\Gamma_{\mathsf{a}\mathsf{a}}$ is an $n\times n$ matrix with components
$({\partial h}/{\partial a_i}|{\partial h}/{\partial a_j})$ $(i,j=1,\ldots,n)$,
$\Gamma_{\mathsf{a}\bxi}$ is an $n\times m$ matrix with components
$({\partial h}/{\partial a_i}|{\partial h}/{\partial\xi_\mathcal{A}})$
$(i=1,\ldots,n,\,\mathcal{A}=1,\ldots,m)$,
and finally $\Gamma_{\bxi\bxi}$ is $m\times m$ matrix with components
$({\partial h}/{\partial\xi_\mathcal{A}}|{\partial h}/{\partial\xi_\mathcal{B}})$
$(\mathcal{A},\mathcal{B}=1,\ldots,m)$.

We introduce two families of the auxiliary $n\times n$ square matrices
$\mathsf{F}_{(\mathcal{A})}$ and $\mathsf{S}_{(\mathcal{AB})}$
($\mathcal{A},\mathcal{B}=1,\ldots,m$),
which depend on the intrinsic parameters $\bxi$ only
(the indexes $\mathcal{A},\mathcal{B}$ within parentheses mean
that they serve here as the matrix labels).
The components of the matrices $\mathsf{F}_{(\mathcal{A})}$
and $\mathsf{S}_{(\mathcal{AB})}$ are defined as follows:
\begin{eqnarray}
\label{Fdef}
F_{(\mathcal{A}){ij}}(\bxi)
&:=& \bigg( h_i(t;\bxi) \bigg|
\frac{\partial h_j(t;\bxi)}{\partial\xi_\mathcal{A}} \bigg),
\quad i,j=1,\ldots,n, \quad \mathcal{A}=1,\ldots,m,
\\\noalign{\smallskip}
\label{Sdef}
S_{(\mathcal{A}\mathcal{B}){ij}}(\bxi)
&:=& \bigg( \frac{\partial h_i(t;\bxi)}{\partial\xi_\mathcal{A}} \bigg|
\frac{\partial h_j(t;\bxi)}{\partial\xi_\mathcal{B}} \bigg),
\quad i,j=1,\ldots,n, \quad \mathcal{A},\mathcal{B}=1,\ldots,m.
\end{eqnarray}
Making use of the definitions~\eqref{eq:M} and \eqref{Fdef}--\eqref{Sdef}
one can write the more explicit form of the matrices
$\Gamma_{\mathsf{a}\mathsf{a}}$, $\Gamma_{\mathsf{a}\bxi}$,
and $\Gamma_{\bxi\bxi}$,
\begin{eqnarray}
\Gamma_{\mathsf{a}\mathsf{a}}(\bxi) &=& \mathsf{M}(\bxi),
\\\noalign{\smallskip}
\Gamma_{\mathsf{a}\bxi}(\mathsf{a},\bxi) &=& \begin{pmatrix}
\mathsf{F}_{(1)}(\bxi) \cdot \mathsf{a} & \cdots & \mathsf{F}_{(m)}(\bxi) \cdot \mathsf{a}
\end{pmatrix},
\\\noalign{\smallskip}
\Gamma_{\bxi\bxi}(\mathsf{a},\bxi) &=& \begin{pmatrix}
\mathsf{a}^\mathsf{T} \cdot \mathsf{S}_{(11)}(\bxi) \cdot \mathsf{a} & \cdots
& \mathsf{a}^\mathsf{T} \cdot \mathsf{S}_{(1m)}(\bxi) \cdot \mathsf{a} \\
\hdotsfor{3} \\\noalign{\smallskip}
\mathsf{a}^\mathsf{T} \cdot \mathsf{S}_{(m1)}(\bxi) \cdot \mathsf{a} & \cdots
& \mathsf{a}^\mathsf{T} \cdot \mathsf{S}_{(mm)}(\bxi) \cdot \mathsf{a}
\end{pmatrix}.
\end{eqnarray}
The notation introduced above means that
the matrix $\Gamma_{\mathsf{a}\bxi}$ can be thought of as a $1\times m$ row matrix
made of $n\times1$ column matrices $\mathsf{F}_{(\mathcal{A})}\cdot\mathsf{a}$.
Thus, the general formula for the component of the matrix $\Gamma_{\mathsf{a}\bxi}$ is
\be
\label{Gaxi2}
\big(\Gamma_{\mathsf{a}\bxi}\big)_{i\mathcal{A}}
= \big(\mathsf{F}_{(\mathcal{A})} \cdot \mathsf{a}\big)_i
= \sum_{j=1}^n F_{(\mathcal{A})ij} \, a_j,
\quad  \mathcal{A}=1,\ldots,m, \quad i=1,\ldots,n.
\ee
The general component of the matrix $\Gamma_{\bxi\bxi}$ is given by
\be
\label{Gxixi2}
\big(\Gamma_{\bxi\bxi}\big)_{\mathcal{A}\mathcal{B}}
= \mathsf{a}^\mathsf{T} \cdot \mathsf{S}_{(\mathcal{A}\mathcal{B})} \cdot \mathsf{a}
= \sum_{i=1}^n \sum_{j=1}^n S_{(\mathcal{A}\mathcal{B})ij} a_i a_j,
\quad  \mathcal{A},\mathcal{B}=1,\ldots,m.
\ee

The \emph{covariance matrix} $\mathsf{C}$, which approximates the expected
covariances of the ML estimators of the parameters $\btheta$, is defined as $\Gamma^{-1}$.
Applying the standard formula for the inverse of a block matrix~\cite{meyer-00}
to Eq.~\eqref{G}, one gets
\be
\label{eq:CC}
\mathsf{C}(\mathsf{a},\bxi) = \begin{pmatrix}
\mathsf{C}_{\mathsf{a}\mathsf{a}}(\mathsf{a},\bxi)
& \mathsf{C}_{\mathsf{a}\bxi}(\mathsf{a},\bxi)
\\\noalign{\smallskip}
\mathsf{C}_{\mathsf{a}\bxi}(\mathsf{a},\bxi)^\mathsf{T}
& \mathsf{C}_{\bxi\bxi}(\mathsf{a},\bxi)
\end{pmatrix},
\ee
where the matrices $\mathsf{C}_{\mathsf{a}\mathsf{a}}$, $\mathsf{C}_{\mathsf{a}\bxi}$,
and $\mathsf{C}_{\bxi\bxi}$ can be expressed in terms of the matrices
$\Gamma_{\mathsf{a}\mathsf{a}}=\mathsf{M}$, $\Gamma_{\mathsf{a}\bxi}$,
and $\Gamma_{\bxi\bxi}$ as follows:
\begin{eqnarray}
\label{Caa}
\mathsf{C}_{\mathsf{a}\mathsf{a}}(\mathsf{a},\bxi) &=& \mathsf{M}(\bxi)^{-1}
+ \mathsf{M}(\bxi)^{-1} \cdot \Gamma_{\mathsf{a}\bxi}(\mathsf{a},\bxi)
\cdot \overline{\Gamma}(\mathsf{a},\bxi)^{-1}
\cdot \Gamma_{\mathsf{a}\bxi}(\mathsf{a},\bxi)^\mathsf{T}
\cdot \mathsf{M}(\bxi)^{-1},
\\\noalign{\smallskip}
\label{Caxi}
\mathsf{C}_{\mathsf{a}\bxi}(\mathsf{a},\bxi)
&=& -\mathsf{M}(\bxi)^{-1} \cdot \Gamma_{\mathsf{a}\bxi}(\mathsf{a},\bxi)
\cdot \overline{\Gamma}(\mathsf{a},\bxi)^{-1},
\\\noalign{\smallskip}
\label{Cxixi}
\mathsf{C}_{\bxi\bxi}(\mathsf{a},\bxi)
&=& \overline{\Gamma}(\mathsf{a},\bxi)^{-1}.
\end{eqnarray}
In Eqs.~\eqref{Caa}\,--\,\eqref{Cxixi} we have introduced the $m\times m$ matrix:
\be
\label{eq:Fproj}
\overline{\Gamma}(\mathsf{a},\bxi) := \Gamma_{\bxi\bxi}(\mathsf{a},\bxi)
- \Gamma_{\mathsf{a}\bxi}(\mathsf{a},\bxi)^\mathsf{T} \cdot \mathsf{M}(\bxi)^{-1}
\cdot \Gamma_{\mathsf{a}\bxi}(\mathsf{a},\bxi).
\ee
We call the matrix $\overline{\Gamma}$
(which is the \textit{Schur complement} of the matrix $\mathsf{M}$)
the \textit{projected Fisher matrix} (onto the space of intrinsic parameters).
Because the matrix $\overline{\Gamma}$ is the inverse
of the intrinsic-parameter submatrix $\mathsf{C}_{\bxi\bxi}$
of the covariance matrix $\mathsf{C}$,
it expresses the information available about the intrinsic parameters
that takes into account the correlations with the extrinsic parameters.
The matrix $\overline{\Gamma}$ is still a function of the putative extrinsic parameters.

We next define the \emph{normalized projected Fisher matrix}
(which is the $m\times m$ square matrix)
\begin{equation}
\label{eq:Fprojn}
\overline{\Gamma}_n(\mathsf{a},\bxi)
:= \frac{\overline{\Gamma}(\mathsf{a},\bxi)}{\rho(\mathsf{a},\bxi)^2},
\end{equation}
where $\rho$ is the signal-to-noise ratio.
Making use of the definition~\eqref{eq:Fproj} and Eqs.~\eqref{Gaxi2}--\eqref{Gxixi2}
we can show that the components of this matrix can be written in the form
\be
\big(\overline{\Gamma}_n(\mathsf{a},\bxi)\big)_{\mathcal{A}\mathcal{B}}
= \frac{\mathsf{a}^\mathsf{T} \cdot \mathsf{A}_{(\mathcal{A}\mathcal{B})}(\bxi) \cdot \mathsf{a}}
{\mathsf{a}^\mathsf{T} \cdot \mathsf{M}(\bxi) \cdot \mathsf{a}}, \quad
\mathcal{A},\mathcal{B}=1,\ldots,m,
\ee
where $\mathsf{A}_{(\mathcal{A}\mathcal{B})}$ is the $n\times n$ matrix defined as
\be
\mathsf{A}_{(\mathcal{A}\mathcal{B})}(\bxi) := S_{(\mathcal{A}\mathcal{B})}(\bxi)
- \mathsf{F}_{(\mathcal{A})}(\bxi)^\mathsf{T}
\cdot \mathsf{M}(\bxi)^{-1}
\cdot \mathsf{F}_{(\mathcal{B})}(\bxi) ,
\quad \mathcal{A},\mathcal{B}=1,\ldots,m.
\ee
From the Rayleigh principle~\cite{meyer-00} it follows
that the minimum value of the component
$(\overline{\Gamma}_n(\mathsf{a},\bxi))_{\mathcal{A}\mathcal{B}}$
is given by the smallest eigenvalue of the matrix
$\mathsf{M}^{-1}\cdot\mathsf{A}_{(\mathcal{A}\mathcal{B})}$. Similarly,
the maximum value of the component
$(\overline{\Gamma}_n(\mathsf{a},\bxi))_{\mathcal{A}\mathcal{B}}$
is given by the largest eigenvalue of that matrix.

Because the trace of a matrix is equal to the sum of its eigenvalues,
the $m\times m$ square matrix $\widetilde{\Gamma}$ with components
\be
\label{eq:rF}
\big(\widetilde{\Gamma}(\bxi)\big)_{\mathcal{A}\mathcal{B}}
:= \frac{1}{n} \, \text{Tr} \Big(
\mathsf{M}(\bxi)^{-1} \cdot \mathsf{A}_{(\mathcal{A}\mathcal{B})}(\bxi) \Big),
\quad \mathcal{A},\mathcal{B}=1,\ldots,m,
\ee
expresses the information available about the intrinsic parameters,
averaged over the possible values of the extrinsic parameters.
Note that the factor $1/n$ is specific to the case of $n$ extrinsic parameters.
We shall call $\widetilde{\Gamma}$ the \textit{reduced Fisher matrix}.
This matrix is a function of the intrinsic parameters alone.
We shall see that the reduced Fisher matrix plays a key role
in the signal processing theory that we present here.
It is used in the calculation of the threshold for statistically significant detection
and in the formula for the number of templates needed to do a given search.

For the case of the signal
\begin{equation}
h(t;A_0,\phi_0,\bxi)
= A_0 \, g(t;\bxi) \cos\left(\phi(t;\bxi)-\phi_0\right),
\end{equation}
the normalized projected Fisher matrix $\overline{\Gamma}_n$ is independent
of the extrinsic parameters $A_0$ and $\phi_0$, and it is equal to the
reduced matrix $\widetilde{\Gamma}$~\cite{owen-96}. The components of
$\widetilde{\Gamma}$ are given by
\begin{equation}
\widetilde{\Gamma}_{\mathcal{A}\mathcal{B}}
= (\Gamma_0)_{\mathcal{A}\mathcal{B}}
- \frac{(\Gamma_0)_{\phi_0\mathcal{A}}(\Gamma_0)_{\phi_0\mathcal{B}}}
{(\Gamma_0)_{\phi_0 \phi_0}},
\end{equation}
where $\Gamma_0$ is the Fisher matrix for the signal
$g(t;\bxi)\cos\left(\phi(t;\bxi)-\phi_0\right)$.


\subsection{False alarm and detection probabilities}


\subsubsection{False alarm and detection probabilities
for known intrinsic parameters}

We first present the false alarm and detection probabilities when the
intrinsic parameters $\bxi$ of the signal are known. In this case the
${\mathcal F}$-statistic is a quadratic form of the random
variables that are correlations of the data. As we assume that
the noise in the data is Gaussian and the correlations are linear
functions of the data, ${\mathcal F}$ is a quadratic form of the
Gaussian random variables. Consequently the ${\mathcal F}$-statistic
has a distribution related to the $\chi^2$ distribution.
One can show (see Section~III~B in~\cite{jaranowski-00})
that for the signal given by Eq.~(\ref{eq:gsig}),
$2{\mathcal F}$ has a $\chi^2$ distribution
with $n$ degrees of freedom when the signal is absent
and noncentral $\chi^2$ distribution with $n$ degrees of freedom
and non-centrality parameter equal to the square of the signal-to-noise ratio
when the signal is present.

As a result the pdfs $p_0$ and $p_1$ of the $\F$-statistic, when the intrinsic
parameters are known and when respectively the signal is absent or
present in the data, are given by
\begin{eqnarray}
  \label{p0}
  p_0({\mathcal F}) &=&
  \frac{{\mathcal F}^{n/2-1}}{(n/2 -1)!}\exp(-{\mathcal F}),
  \\[1ex]
  \label{p1}
  p_1({\rho,\mathcal F}) &=&
  \frac{(2{\mathcal F})^{(n/2 -1)/2}}{\rho^{n/2-1}} I_{n/2-1}
  \left(\rho\sqrt{2 {\mathcal F}}\right)
  \exp\left(-{\mathcal F}-\frac{1}{2}\rho^2\right),
\end{eqnarray}%
where $n$ is the number of degrees of freedom of $\chi^2$ distribution
and $I_{n/2-1}$ is the modified Bessel function of the first kind and order $n/2-1$.
The \emph{false alarm probability} $P_\mathrm{F}$ is the probability
that $\F$ exceeds a certain threshold $\Fo$ when there is no signal.
In our case we have
\begin{equation}
  \label{PF}
  P_\mathrm{F}(\Fo) := \int_{\Fo}^\infty \!\! p_0(\F)\,\md\F =
  \exp(-\Fo) \sum^{n/2-1}_{k=0}\frac{\Fo^k}{k!}.
\end{equation}
The \emph{probability of detection} $P_\mathrm{D}$ is the probability that
$\F$ exceeds the threshold $\Fo$ when a signal is present
and the signal-to-noise ratio is equal to $\rho$:
\begin{equation}
  \label{PD}
  P_\mathrm{D}(\rho,\Fo) := \int^{\infty}_{\Fo} \!\! p_1(\rho,\F)\,\md\F.
\end{equation}
The integral in the above formula can be expressed in terms of the
generalized Marcum $Q$-function~\cite{marcum-50, helstrom-95},
$P_\mathrm{D}(\rho,\Fo)=Q(\rho,\sqrt{2\Fo})$. We see that when the
noise in the detector is Gaussian and the intrinsic parameters are
known, the probability of detection of the signal depends on a single
quantity: the optimal signal-to-noise ratio $\rho$.


\subsubsection{False alarm probability for unknown intrinsic parameters}
\label{sec:FA}

Next we return to the case in which the intrinsic parameters $\bxi$ are not known.
Then the statistic $\F[x;\bxi]$ given by Eq.~(\ref{eq:FS})
is a certain multiparameter random process
called the \emph{random field} (see monographs~\cite{adler-81,AdlerTaylor2007}
for a comprehensive discussion of random fields).
If the vector $\bxi$ has one
component the random field is simply a random process.
For random fields we define the \emph{autocovariance function} ${\cal C}$
just in the same way as we define such a function for a random process:
\begin{equation}
  {\cal C}({\mathbold\xi},{\mathbold\xi'}) :=
  \mathrm{E}_0\big[{\mathcal F}[x;\bxi]{\mathcal F}[x;\bxi']\big]
- \mathrm{E}_0\big[{\mathcal F}[x;\bxi]\big]
  \mathrm{E}_0\big[{\mathcal F}[x;\bxi']\big],
\end{equation}
where ${\mathbold\xi}$ and ${\mathbold\xi'}$ are two values of the
intrinsic parameter set, and $\mathrm{E}_0$ is the expectation value
when the signal is absent. One can show that for the
signal~(\ref{eq:gsig}) the autocovariance function ${\cal C}$ is given
by
\begin{equation}
\label{eq:autocov}
{\cal C}(\bxi,\bxi') = \frac{1}{2} \, \text{Tr} \Big(
\mathsf{Q}(\bxi,\bxi') \cdot \mathsf{M}(\bxi')^{-1}
\cdot \mathsf{Q}(\bxi,\bxi')^\mathsf{T} \cdot \mathsf{M}(\bxi)^{-1} \Big),
\end{equation}
where $\mathsf{Q}$ is an $n\times n$ matrix with components
\begin{equation}
\label{eq:Q}
\mathsf{Q}(\bxi,\bxi')_{ij} :=
\left(h_i(t;\bxi)|h_j(t;\bxi')\right),
\quad i,j=1,\ldots,n.
\end{equation}
Obviously $\mathsf{Q}(\bxi,\bxi)=\mathsf{M}(\bxi)$,
therefore $\mathcal{C}(\bxi,\bxi)=n/2$.

One can estimate the false alarm probability in the following
way~\cite{jaranowski-98}. The autocovariance function ${\cal C}$
tends to zero as the displacement
$\Delta{\mathbold\xi}={\mathbold\xi}'-{\mathbold\xi}$ increases
(it is maximal for $\Delta{\mathbold\xi}={\mathbold 0}$). Thus we
can divide the parameter space into \emph{elementary cells} such
that in each cell the autocovariance function ${\cal C}$ is
appreciably different from zero. The realizations of the random
field within a cell will be correlated (dependent), whereas
realizations of the random field within each cell and outside of the
cell are almost uncorrelated (independent). Thus, the number of
cells covering the parameter space gives an estimate of the number
of independent realizations of the random field.

We choose the elementary cell with its origin at the point $\bxi$
to be a compact region with boundary defined by the requirement
that the autocovariance ${\cal C}(\bxi,\bxi')$ between the origin $\bxi$
and any point $\bxi'$ at the cell's boundary equals half of its maximum value,
i.e., $\mathcal{C}(\bxi,\bxi)/2$. Thus, the elementary cell is defined by
the inequality
\begin{equation}
\label{ecell1}
\mathcal{C}(\bxi,\bxi') \le \frac{1}{2}\mathcal{C}(\bxi,\bxi) = \frac{n}{4},
\end{equation}
with $\mathbold\xi$ at the cell's center and ${\mathbold\xi}'$ on the cell's boundary.

To estimate the number of cells we perform the Taylor
expansion of the autocovariance function up to the second-order terms:
\begin{equation}
\label{ecell2}
\mathcal{C}(\bxi,\bxi') \cong \frac{n}{2}
+ \sum_{\mathcal{A}=1}^m \frac{\partial{\cal C}(\bxi,\bxi')}
{\partial \xi'_\mathcal{A}}\bigg|_{\mathboldsmall{\xi}'=\mathboldsmall{\xi}}
\!\!\Delta\xi_\mathcal{A} + \frac{1}{2} \sum_{\mathcal{A},\mathcal{B}=1}^m
\frac{\partial^2 {\cal C}(\bxi,\bxi')}
{\partial\xi'_\mathcal{A}\,\partial\xi'_\mathcal{B}}
\bigg|_{\mathboldsmall{\xi}'=\mathboldsmall{\xi}}
\!\!\Delta\xi_\mathcal{A}\,\Delta\xi_\mathcal{B}.
\end{equation}
As ${\cal C}$ attains its maximum
value when $\mathbold{\xi}-\mathbold{\xi}'=\mathbold{0}$, we have
\begin{equation}
\label{ecell3}
\frac{\partial{\cal C}(\bxi,\bxi')}{\partial \xi'_\mathcal{A}}
\bigg|_{\mathboldsmall{\xi}'=\mathboldsmall{\xi}} \!\!\! = 0,
\quad \mathcal{A}=1,\ldots,m.
\end{equation}
Let us introduce the symmetric matrix $\mathsf{G}$ with components
\begin{equation}
\label{eq:G}
G_{\mathcal{A}\mathcal{B}}(\bxi)
:= -\frac{1}{2\,\mathcal{C}(\bxi,\bxi)}
\frac{\partial^2 {\cal C}(\bxi,\bxi')}
{\partial\xi'_\mathcal{A}\,\partial\xi'_\mathcal{B}}
\bigg|_{\mathboldsmall{\xi'}=\mathboldsmall{\xi}},
\quad \mathcal{A},\mathcal{B}=1,\ldots,m.
\end{equation}
Then, the inequality~\eqref{ecell1} for the elementary cell
can approximately be written as
\begin{equation}
\label{eq:GV}
\sum_{\mathcal{A},\mathcal{B}=1}^m
G_{\mathcal{A}\mathcal{B}}(\bxi)\,
\Delta\xi_\mathcal{A}\,\Delta\xi_\mathcal{B}
\le \frac{1}{2}.
\end{equation}
It is interesting to find a relation between the matrix $G$ and the
Fisher matrix. One can show (see~\cite{krolak-04}, Appendix~B) that
the matrix $G$ is precisely equal to the reduced Fisher matrix
$\tilde{\Gamma}$ given by Eq.~(\ref{eq:rF}).

If the components of the matrix $\mathsf{G}$ are constant
(i.e., they are independent of the values of the intrinsic parameters $\bxi$ of the signal),
the above equation defines a hyperellipsoid in $m$-dimensional
($m$ is the number of the intrinsic parameters) Euclidean space $\mathbb{R}^m$.
The $m$-dimensional Euclidean volume $V_\mathrm{cell}$
of the elementary cell given by Eq.~(\ref{eq:GV}) equals
\begin{equation}
\label{eq:vc}
V_\mathrm{cell} = \frac{(\pi/2)^{m/2}}{\Gamma(m/2+1)\sqrt{\det\mathsf{G}}},
\end{equation}
where $\Gamma$ denotes the Gamma function. We estimate the number
$\nc$ of elementary cells by dividing the total Euclidean
volume $V$ of the $m$-dimensional intrinsic parameter space by the volume
$V_\mathrm{cell}$ of one elementary cell, i.e., we have
\begin{equation}
\label{NC}
\nc = \frac{V}{V_{\mathrm{cell}}}.
\end{equation}
The components of the matrix $\mathsf{G}$ are constant for the signal
$h(t;A_0,\phi_0,\bxi)=A_0\cos\left(\phi(t;\bxi)-\phi_0\right)$,
provided the phase $\phi(t;\bxi)$ is a \emph{linear} function of the
intrinsic parameters $\bxi$.

To estimate the number of cells in the case when the components of
the matrix $\mathsf{G}$ are not constant, i.e., when they depend on the
values of the intrinsic parameters $\bxi$, one replaces Eq.~(\ref{NC}) by
\begin{equation}
\label{eq:NC}
\nc = \frac{\Gamma(m/2+1)}{(\pi/2)^{m/2}}
\int_V\sqrt{\det\mathsf{G}(\bxi)}\,\md V.
\end{equation}
This formula can be thought of as interpreting the matrix $\mathsf{G}$ as the
metric on the parameter space. This interpretation appeared for the
first time in the context of gravitational-wave data analysis in the
work by Owen~\cite{owen-96}, where an analogous integral formula was
proposed for the number of templates needed to perform a search for
gravitational-wave signals from coalescing binaries.

The concept of number of cells was introduced in~\cite{jaranowski-98}
and it is a generalization of the idea of an effective number of
samples introduced in~\cite{dhurandhar2-94} for the case of a
coalescing binary signal.

We approximate the pdf of the $\F$-statistic in each cell
by the pdf $p_0(\F)$ of the $\F$-statistic when the parameters are known
[it is given by Eq.~(\ref{p0})]. The values
of the $\F$-statistic in each cell can be considered as independent
random variables. The probability that $\F$ does not exceed the
threshold $\Fo$ in a given cell is $1-P_\mathrm{F}(\Fo)$, where
$P_\mathrm{F}(\Fo)$ is given by Eq.~(\ref{PF}).
Consequently the probability that $\F$ does not exceed the threshold $\Fo$
in \emph{all} the $\nc$ cells is
$[1-P_\mathrm{F}(\Fo)]^\nc$.
Thus, the probability $P^T_\mathrm{F}$ that $\F$ exceeds $\Fo$
in \emph{one or more} cells is given by
\begin{equation}
\label{FP}
P^T_\mathrm{F}(\Fo) = 1 - [1 - P_\mathrm{F}(\Fo)]^{\nc}.
\end{equation}
By definition, this is the false alarm probability when the phase
parameters are unknown.
The number of false alarms $N_\mathrm{F}$ is given by
\begin{equation}
\label{NF}
N_\mathrm{F} = \nc P^T_\mathrm{F}(\Fo).
\end{equation}
A different approach to the calculation of the number of false alarms
using the Euler characteristic of level crossings of a random field
is described in~\cite{jaranowski-00}.

It was shown (see~\cite{pinto-04}) that for any finite $\Fo$ and
$\nc$, Eq.~(\ref{FP}) provides an upper bound for the
false alarm probability. Also in~\cite{pinto-04} a tighter upper bound
for the false alarm probability was derived by modifying a formula
obtained by Mohanty~\cite{mohanty-98}. The formula amounts essentially
to introducing a suitable coefficient multiplying the number $\nc$ of cells.


\subsubsection{Detection probability for unknown intrinsic parameters}

When the signal is present in the data a precise calculation of the pdf of
the $\F$-statistic is very difficult because the presence of the signal
makes the data's random process non-stationary. As a first
approximation we can estimate the probability of detection of the
signal when the intrinsic parameters are unknown by the probability of
detection when these parameters are known [it is given by
  Eq.~(\ref{PD})]. This approximation assumes that when the signal is
present the true values of the intrinsic parameters fall within the
cell where the $\F$-statistic has a maximum. This approximation will
be the better the higher the signal-to-noise ratio $\rho$ is.


\subsection{Number of templates}

To search for gravitational-wave signals we evaluate the
$\F$-statistic on a grid in parameter space. The grid has to be
sufficiently fine such that the loss of signals is minimized. In order
to estimate the number of points of the grid, or in other words the
number of templates that we need to search for a signal, the natural
quantity to study is the expectation value of the $\F$-statistic when
the signal is present.

Thus, we assume that the data $x$ contains the gravitational-wave signal
$h(t;\btheta)$ defined in Eq.~\eqref{gsig2}, so $x(t;\btheta)=h(t;\btheta)+n(t)$.
The parameters $\btheta=(\mathsf{a},\bxi)$ of the signal consist of
extrinsic parameters $\mathsf{a}$ and intrinsic parameters $\bxi$.
The data $x$ will be correlated with the filters $h_i(t;\bxi')$ ($i=1,\ldots,n$)
parameterized by the values $\bxi'$ of the intrinsic parameters.
The $\F$-statistic can thus be written in the form
[see Eq.~\eqref{eq:FS}]
\be
\label{not01}
\F[x(t;\mathsf{a},\bxi);\bxi']
= \frac{1}{2}\, \mathsf{N}[x(t;\mathsf{a},\bxi);\bxi']^\mathsf{T}
\cdot \mathsf{M}(\bxi')^{-1} \cdot \mathsf{N}[x(t;\mathsf{a},\bxi);\bxi'],
\ee
where the matrices $\mathsf{M}$ and $\mathsf{N}$ are defined in Eqs.~\eqref{eq:M}.
The expectation value of the $\F$-statistic~\eqref{not01} is
\be
\label{evofF}
\mathrm{E}\big[\F[x(t;\mathsf{a},\bxi);\bxi']\big] = \frac{1}{2} \left( n +
\mathsf{a}^\mathsf{T} \cdot \mathsf{Q}(\bxi,\bxi') \cdot \mathsf{M}(\bxi')^{-1}
\cdot \mathsf{Q}(\bxi,\bxi')^\mathsf{T} \cdot \mathsf{a} \right),
\ee
where the matrix $\mathsf{Q}$ is defined in Eq.\ (\ref{eq:Q}).
Let us rewrite the expectation value~(\ref{evofF}) in the following form,
\begin{equation}
\mathrm{E}\big[\F[x(t;\mathsf{a},\bxi);\bxi']\big]
= \frac{1}{2} \left( n + \rho(\mathsf{a},\bxi)^2
{\cal C}_\mathrm{n}(\mathsf{a},\bxi,\bxi') \right),
\end{equation}
where $\rho$ is the signal-to-noise ratio and where we have introduced
the \emph{normalized correlation function} ${\cal C}_\mathrm{n}$,
\begin{equation}
{\cal C}_\mathrm{n}(\mathsf{a},\bxi,\bxi')
:= \frac{\mathsf{a}^\mathsf{T} \cdot \mathsf{Q}(\bxi,\bxi')
\cdot \mathsf{M}(\bxi')^{-1} \cdot \mathsf{Q}(\bxi,\bxi')^\mathsf{T} \cdot \mathsf{a}}
{\mathsf{a}^\mathsf{T}\cdot\mathsf{M}(\bxi)\cdot\mathsf{a}}.
\end{equation}
From the Rayleigh principle~\cite{meyer-00} it follows that
the minimum value of the normalized correlation function
is equal to the smallest eigenvalue of the matrix
$\mathsf{M}(\bxi)^{-1}\cdot\mathsf{Q}(\bxi,\bxi')
\cdot \mathsf{M}(\bxi')^{-1} \cdot \mathsf{Q}(\bxi,\bxi')^\mathsf{T}$,
whereas the maximum value is given by its largest eigenvalue.
We define the \emph{reduced correlation function} ${\cal C}$ as
\begin{equation}
\label{eq:CG}
{\cal C}(\bxi,\bxi') := \frac{1}{2} \tr \big(
\mathsf{M}(\bxi)^{-1}\cdot\mathsf{Q}(\bxi,\bxi')
\cdot \mathsf{M}(\bxi')^{-1} \cdot \mathsf{Q}(\bxi,\bxi')^\mathsf{T} \big).
\end{equation}
As the trace of a matrix equals the sum of its eigenvalues, the
reduced correlation function ${\cal C}$ is equal to the average of the
eigenvalues of the normalized correlation function
${\cal C}_\mathrm{n}$. In this sense we can think of the reduced
correlation function as an ``average'' of the normalized correlation
function. The advantage of the reduced correlation function is that it
depends only on the intrinsic parameters $\mathbold\xi$, and thus
is suitable for studying the number of grid points on which the
$\F$-statistic needs to be evaluated. We also note that the normalized
correlation function ${\cal C}$ precisely coincides with the
autocovariance function ${\cal C}$ of the $\F$-statistic given by
Eq.~(\ref{eq:autocov}).

As in the calculation of the number of cells in order to estimate
the number of templates we perform a Taylor expansion of ${\cal C}$ up
to second order terms around the true values of the parameters,
and we obtain an equation analogous to Eq.~(\ref{eq:GV}),
\begin{equation}
\label{eq:e}
\sum_{\mathcal{A},\mathcal{B}=1}^m \mathsf{G}_{\mathcal{A}\mathcal{B}}
\, \Delta\xi_\mathcal{A} \, \Delta\xi_\mathcal{B} = 1 - C_0,
\end{equation}
where $\mathsf{G}$ is given by Eq.~(\ref{eq:G}).
By arguments identical to those in deriving the formula for the number of cells
we arrive at the following formula for the number of templates:
\begin{equation}
  \label{eq:nt}
  N_\mathrm{t} = \frac{1}{(1 - C_0)^{m/2}} \frac{\Gamma(m/2+1)}{\pi^{m/2}}
  \int_V\sqrt{\det\mathsf{G}(\bxi)}\,\md V.
\end{equation}
When $C_0=1/2$ the above formula coincides with the formula for the
number $\nc$ of cells, Eq.~(\ref{eq:NC}). Here we would
like to place the templates sufficiently closely so that the loss of
signals is minimized. Thus $1-C_0$ needs to be chosen sufficiently
small. The formula~(\ref{eq:nt}) for the number of templates assumes
that the templates are placed in the centers of hyperspheres and that
the hyperspheres fill the parameter space without holes. In order to
have a tiling of the parameter space without holes we can place the
templates in the centers of hypercubes, which are inscribed in the
hyperspheres. Then the formula for the number of templates reads
\begin{equation}
\label{eq:ntm}
N_\mathrm{t} = \frac{1}{(1 - C_0)^{m/2}} \frac{m^{m/2}}{2^m}
\int_V \sqrt{\det\mathsf{G}(\bxi)}\,\md V.
\end{equation}

For the case of the signal given by Eq.~(\ref{eq:chsig}) our
formula for the number of templates is equivalent to the original formula
derived by Owen~\cite{owen-96}. Owen~\cite{owen-96} has also
introduced a geometric approach to the problem of template placement
involving the identification of the Fisher matrix with a metric on the
parameter space. An early study of the template placement for the case
of coalescing binaries can be found in~\cite{sathya-91, dhurandhar-94, bala-95}.
Applications of the geometric approach of Owen to the case
of spinning neutron stars and supernova bursts are given
in~\cite{brady-98, arnaud-03}.

\subsubsection{Covering problem}
\label{sec:covering-problem}

The problem of how to cover the parameter space with the smallest
possible number of templates, such that no point in the parameter
space lies further away from a grid point than a certain distance,
is known in mathematical literature as the \emph{covering problem}~\cite{conway-99}.
This was first studied in the context of gravitational-wave data analysis
by Prix~\cite{Prix2007b}.
The maximum distance of any point to the
next grid point is called the \emph{covering radius} $R$. An important
class of coverings are \emph{lattice coverings}. We define a lattice
in $m$-dimensional Euclidean space ${\mathbb R}^m$ to be the set of
points including 0 such that if $u$ and $v$ are lattice points, then
also $u + v$ and $u - v$ are lattice points. The basic building block
of a lattice is called the \emph{fundamental region}. A lattice
covering is a covering of ${\mathbb R}^m$ by spheres of covering
radius $R$, where the centers of the spheres form a lattice. The most
important quantity of a covering is its \emph{thickness} $\Theta$
defined as
\begin{equation}
  \Theta := \frac{\mbox{volume of one }m\mbox{-dimensional sphere}}
  {\mbox{volume of the fundamental region}}.
\end{equation}
In the case of a two-dimensional Euclidean space the best covering is
the hexagonal covering and its thickness $\simeq 1.21$. For dimensions
higher than 2 the best covering is not known. However, we know the best
lattice covering for dimensions $m\leq23$. These are \emph{$A^*_m$ lattices}, which have thicknesses $\Theta_{A^*_m}$ equal to
\begin{equation}
  \Theta_{A^*_m} = V_m \sqrt{m+1}
  \left(\frac{m(m+2)}{12(m+1)}\right)^{m/2},
\end{equation}
where $V_m$ is the volume of the $m$-dimensional sphere of unit
radius. The advantage of an $A^{*}_m$ lattice over the
hypercubic lattice grows exponentially with the number of dimensions.

For the case of gravitational-wave signals from spinning neutron stars
a 3-dimensional grid was constructed~\cite{astone-02}.
It consists of prisms with hexagonal bases. Its thickness is around 1.84,
which is much better than the cubic grid with a thickness of approximately 2.72.
It is worse than the best 3-dimensional lattice covering,
which has a thickness of around 1.46.

In~\cite{astone-10} a grid was constructed in the 4-dimensional
parameter space spanned by frequency, frequency derivative, and sky position of the source,
for the case of an almost monochromatic gravitational-wave signal
originating from a spinning neutron star.
The starting point of the construction was an $A^{*}_4$ lattice of
thickness $\simeq 1.77$. The grid was then constrained so that the nodes
of the grid coincide with Fourier frequencies. This allowed the use
of a fast Fourier transform (FFT) to evaluate the maximum-likelihood $\F$-statistic efficiently
(see Section~\ref{ssec:FFT}).
The resulting lattice is only 20\% thicker than the optimal $A^{*}_4$ lattice.

Efficient 2-dimensional banks of templates suitable for directed searches
(in which one assumes that the position of the gravitational-wave source in the sky is known,
but one does not assume that the wave's frequency and its derivative are \emph{a priori} known)
were constructed in~\cite{PisarskiJaranowskiPietka2011}.
All grids found in~\cite{PisarskiJaranowskiPietka2011} enable usage of the FFT
algorithm in the computation of the $\F$-statistic; they have thicknesses 0.1\,--\,16\% larger
than the thickness of the optimal 2-dimensional hexagonal covering.
In the construction of grids the dependence on the choice of the position of the
observational interval with respect to the origin of time axis was employed.
Also the usage of the FFT algorithms with nonstandard frequency resolutions achieved by zero padding
or folding the data was discussed.

The above template placement constructions are based on a Fisher matrix
with constant coefficients, i.e., they assume that the parameter manifold is flat.
The generalization to curved Riemannian parameter manifolds is difficult.
An interesting idea to overcome this problem is to use stochastic
template banks where a grid in the parameter space is randomly generated
by some algorithm~\cite{Messenger09,Harry09,MancaVallisneri10,Rover2010}.


\subsection{Suboptimal filtering}

To extract gravitational-wave signals from the detector's noise
one very often uses filters that are not optimal.
We may have to choose an approximate, suboptimal filter
because we do not know the exact form of the signal (this is almost
always the case in practice) or in order to reduce the computational
cost and to simplify the analysis.
In the case of the signal of the form given in Eq.~\eqref{gsig2}
the most natural and simplest way to proceed
is to use as detection statistic the $\F$-statistic
where the filters $h'_k(t;\bzeta)$ ($k=1,\ldots,n$) are the approximate ones
instead of the optimal ones $h_k(t;\bxi)$ ($k=1,\ldots,n$) matched to the signal.
In general the functions $h'_k(t;\bzeta)$ will be different from the functions
$h_k(t;\bxi)$ used in optimal filtering, and also the set of
parameters $\bzeta$ will be different from the set of
parameters $\bxi$ in optimal filters. We call this procedure
the \emph{suboptimal filtering} and we denote the \emph{suboptimal statistic} by $\Fs$.
It is defined as [see Eq.~\eqref{eq:FS}]
\be
\label{subof1}
\Fs[x;\bzeta] := \frac{1}{2}\,
\mathsf{N}_\text{s}[x;\bzeta]^\mathsf{T}
\cdot \mathsf{M}_\text{s}(\bzeta)^{-1}
\cdot \mathsf{N}_\text{s}[x;\bzeta],
\ee
where the data-dependent $n\times1$ column matrix $\mathsf{N}_\text{s}$
and the square $n\times n$ matrix $\mathsf{M}_\text{s}$ have components
[see Eq.~\eqref{eq:M}]
\be
\label{subof2}
N_{\text{s}\,i}[x; \bzeta] := (x(t)|h'_i(t;\bzeta)),  \quad
M_{\text{s}\,ij}(\bzeta) := (h'_i(t;\bzeta)|h'_j(t;\bzeta)),
\quad  i,j = 1,\ldots,n.
\ee

We need a measure of how well a given suboptimal filter performs.
To find such a measure we calculate the expectation value of the
suboptimal statistic $\Fs$ in the case where the data contains
the gravitational-wave signal,
i.e., when $x(t;\mathsf{a},\bxi)=n(t)+h(t;\mathsf{a},\bxi)$. We get
\begin{equation}
\label{subof3}
\mathrm{E}\big[\Fs[x(t;\mathsf{a},\bxi);\bzeta]\big]
= \frac{1}{2} \big( n + \mathsf{a}^\mathsf{T}
\cdot \mathsf{Q}_\text{s}(\bxi,\bzeta)
\cdot \mathsf{M}_\text{s}(\bzeta)^{-1}
\cdot \mathsf{Q}_\text{s}(\bxi,\bzeta)^\mathsf{T}
\cdot \mathsf{a} \big),
\end{equation}
where we have introduced the matrix $\mathsf{Q}_\text{s}$ with components
\begin{equation}
\label{eq:MQs}
Q_{\text{s}\,ij}(\bxi,\bzeta) := \big(h_i(t;\bxi)|h'_j(t;\bzeta)\big),
\quad i,j=1,\ldots,n.
\end{equation}
Let us rewrite the expectation value~\eqref{subof3}
in the following form,
\begin{equation}
\mathrm{E}\big[\Fs[x(t;\mathsf{a},\bxi);\bzeta]\big]
= \frac{1}{2} \left( n + \rho(\mathsf{a},\bxi)^2
\frac{\mathsf{a}^\mathsf{T}
\cdot \mathsf{Q}_\text{s}(\bxi,\bzeta)
\cdot \mathsf{M}_\text{s}(\bzeta)^{-1}
\cdot \mathsf{Q}_\text{s}(\bxi,\bzeta)^\mathsf{T}
\cdot \mathsf{a}}
{\mathsf{a}^\mathsf{T} \cdot \mathsf{M}(\bxi) \cdot \mathsf{a}} \right),
\end{equation}
where $\rho$ is the optimal signal-to-noise ratio
[given in Eq.~\eqref{snr2}].
This expectation value reaches its maximum equal to
$(n+\rho^2)/2$ when the filter is perfectly matched to the signal.
Therefore, a natural measure of the performance of a suboptimal filter
is the quantity FF defined by
\begin{equation}
\label{ffg}
\text{FF}(\bxi) := \max_{(\mathsf{a},\bzeta)} \sqrt{
\frac{\mathsf{a}^\mathsf{T}
\cdot \mathsf{Q}_\text{s}(\bxi,\bzeta)
\cdot \mathsf{M}_\text{s}(\bzeta)^{-1}
\cdot \mathsf{Q}_\text{s}(\bxi,\bzeta)^\mathsf{T}
\cdot \mathsf{a}}
{\mathsf{a}^\mathsf{T} \cdot \mathsf{M}(\bxi) \cdot \mathsf{a}} }.
\end{equation}
We call the quantity FF the \emph{generalized fitting factor}.
From the Rayleigh principle, it follows
that the generalized fitting factor
is the maximum of the largest eigenvalue of the matrix
$\mathsf{M}(\bxi)^{-1}\cdot\mathsf{Q}_\text{s}(\bxi,\bzeta)
\cdot \mathsf{M}_\text{s}(\bzeta)^{-1}
\cdot \mathsf{Q}_\text{s}(\bxi,\bzeta)^\mathsf{T}$
over the intrinsic parameters of the signal.

In the case of a gravitational-wave signal given by
\begin{equation}
\label{sig0}
s(t;A_0,\bxi) = A_0 \, h(t;\bxi),
\end{equation}
the generalized fitting factor defined above reduces to the fitting
factor introduced by Apostolatos~\cite{apostolatos-95}:
\begin{equation}
  \label{ffa}
\text{FF}(\bxi) = \max_{\bzeta}
  \frac{\left(h(t;\bxi)\vert h'(t;\bzeta)\right)}
  {\sqrt{\left(h(t;\bxi)\vert h(t;\bxi)\right)}
  \sqrt{\left(h'(t;\bzeta)\vert h'(t;\bzeta)\right)}}.
\end{equation}
The fitting factor is the ratio of the maximal signal-to-noise ratio
that can be achieved with suboptimal filtering to the signal-to-noise
ratio obtained when we use a perfectly matched, optimal filter. We
note that for the signal given by Eq.~(\ref{sig0}), FF is
independent of the value of the amplitude $A_0$.

For the case of a signal of the form
\begin{equation}
\label{sig1}
s(t;A_0,\phi_0,\bxi)
= A_0 \cos\left(\phi(t;\bxi)+\phi_0\right),
\end{equation}
where $\phi_0$ is a constant phase, the maximum over $\phi_0$ in
Eq.~(\ref{ffa}) can be obtained analytically. Moreover, assuming
that over the bandwidth of the signal the spectral density of the
noise is constant and that over the observation time
$\cos\phi(t;{\mathbold\xi})$ oscillates rapidly, the
fitting factor is approximately given by
\begin{equation}
\text{FF}(\bxi) \cong \max_{\bzeta}
\left[ \left(\int^{T_0}_0\!\!\!
\cos\big(\phi(t;\bxi)-\phi'(t;\bzeta)\big)\md t\right)^2 \!\!+
\left(\int^{T_0}_0\!\!\!
\sin\big(\phi(t;\bxi)-\phi'(t;\bzeta)\big)\md t\right)^2 \right]^{1/2}.
\end{equation}

In designing suboptimal filters one faces the issue of how small a
fitting factor one can accept. A popular rule of thumb is accepting
$\mathrm{FF}=0.97$. Assuming that the amplitude of the signal and
consequently the signal-to-noise ratio decreases inversely
proportionally to the distance from the source this corresponds to 10\%
loss of the signals that would be detected by a matched filter.

Proposals for good suboptimal (search) templates for the case of
coalescing binaries are given in~\cite{buonanno-03,tanaka-00} and for
the case-spinning neutron stars in~\cite{jaranowski-00,astone-02}.


\subsection[Algorithms to calculate the ${\F}$-statistic]
          {Algorithms to calculate the \boldmath ${\F}$-statistic}
\label{sec:alg}


\subsubsection{The two-step procedure}

In order to detect signals we search for threshold crossings of the
$\F$-statistic over the intrinsic parameter space. Once we have a
threshold crossing we need to find the precise location of the maximum
of $\F$ in order to estimate accurately the parameters of the
signal. A satisfactory procedure is the two-step procedure. The first
step is a \emph{coarse search} where we evaluate $\F$ on a coarse grid
in parameter space and locate threshold crossings. The second step,
called a \emph{fine search}, is a refinement around the region of
parameter space where the maximum identified by the coarse search is
located.

There are two methods to perform the fine search. One is to refine the
grid around the threshold crossing found by the coarse
search~\cite{mohanty-96, mohanty-98, tanaka-00, sengupta-03}, and the
other is to use an optimization routine to find the maximum of
$\F$~\cite{jaranowski-00, krolak-04}. As initial values to the
optimization routine we input the values of the parameters found by
the coarse search. There are many maximization algorithms
available. One useful method is the Nelder--Mead
algorithm~\cite{lagarias-98}, which does not require computation of the
derivatives of the function being maximized.


\subsubsection[Evaluation of the $\F$-statistic]
             {Evaluation of the \boldmath $\F$-statistic}
\label{ssec:FFT}

Usually the grid in parameter space is very large and it is important
to calculate the optimum statistic as efficiently as possible. In
special cases the $\F$-statistic given by Eq.~(\ref{eq:FS}) can
be further simplified. For example, in the case of coalescing binaries
$\F$ can be expressed in terms of convolutions that depend on the
difference between the time-of-arrival (TOA) of the signal and the TOA
parameter of the filter. Such convolutions can be efficiently computed
using FFTs. For continuous sources, like
gravitational waves from rotating neutron stars observed by
ground-based detectors~\cite{jaranowski-00} or gravitational waves
form stellar mass binaries observed by space-borne
detectors~\cite{krolak-04}, the detection statistic $\F$ involves
integrals of the general form
\begin{equation}
  \int^{T_0}_0 \!\!\! x(t) \, m(t;\omega,\tilde{\bxi}) \,
  \exp\big(i\omega\phi_{\mathrm{mod}}(t;\tilde{\bxi})\big)
  \exp(i\omega t)\,\md t,
  \label{eq:fstat}
\end{equation}
where $\tilde{\bxi}$ are the intrinsic parameters excluding the
frequency parameter $\omega$, $m$ is the amplitude modulation function,
and $\omega \phi_{\mathrm{mod}}$ the phase modulation function. The
amplitude modulation function is slowly varying compared to the
exponential terms in the integral~(\ref{eq:fstat}). We see that the
integral~(\ref{eq:fstat}) can be interpreted as a Fourier transform
(and computed efficiently with an FFT), if $\phi_\mathrm{mod}=0$ and
if $m$ does not depend on the frequency $\omega$. In the
long-wavelength approximation the amplitude function $m$ does not
depend on the frequency. In this case, Eq.~(\ref{eq:fstat}) can be
converted to a Fourier transform by introducing a new time variable
$t_\mathrm{b}$~\cite{schutz-91},
\begin{equation}
  \label{eq:Bt}
  t_\mathrm{b}(t;\tilde{\bxi}) := t + \phi_{\mathrm{mod}}(t;\tilde{\bxi}).
\end{equation}
Thus, in order to compute the integral~(\ref{eq:fstat}), for each set
of the intrinsic parameters $\tilde{\bxi}$ we multiply the data by
the amplitude modulation function $m$, resample according to
Eq.~(\ref{eq:Bt}), and perform the FFT. In the case of LISA
detector data when the amplitude modulation $m$ depends on frequency
we can divide the data into several band-passed data sets, choosing
the bandwidth for each set to be sufficiently small so that the change of
$m\exp(i\omega\phi_{\mathrm{mod}})$ is small over the band. In the
integral~(\ref{eq:fstat}) we can then use as the value of the
frequency in the amplitude and phase modulation function the maximum
frequency of the band of the signal (see~\cite{krolak-04} for details).


\subsection{Accuracy of parameter estimation}

\subsubsection{Fisher-matrix-based assessments}

Fisher matrix has been extensively used to assess the accuracy
of estimation of astrophysically-interesting parameters
of different gravitational-wave signals.
For ground-based interferometric detectors, the
first calculations of the Fisher matrix concerned gravitational-wave signals
from inspiralling compact binaries (made of neutron stars or black holes)
in the leading-order quadrupole approximation~\cite{finn-93,krolak-93,jaranowski-94}
and from quasi-normal modes of Kerr black hole~\cite{finn-92}.

Cutler and Flanagan~\cite{cutler-94} initiated the study of the implications
of the higher-order post-Newtonian (PN) phasing formula
as applied to the parameter estimation of inspiralling binary signals.
They used the 1.5PN phasing formula to investigate
the problem of parameter estimation, both for spinning and
non-spinning binaries, and examined the effect of the spin-orbit
coupling on the estimation of parameters.
The effect of the 2PN phasing formula was analyzed independently
by Poisson and Will~\cite{poisson-95}
and Kr\'olak, Kokkotas and Sch\"afer~\cite{krolak-95}.
In both cases the focus was to understand
the leading-order spin-spin coupling term appearing at the 2PN level
when the spins were aligned perpendicularly to the orbital plane.
Compared to~\cite{krolak-95}, \cite{poisson-95} also included
\emph{a priori} information about the magnitude of the spin parameters,
which then leads to a reduction in the rms errors
in the estimation of mass parameters.
The case of a 3.5PN phasing formula was studied in detail by
Arun et~al.~\cite{arun-05}. Inclusion of 3.5PN effects leads to an
improved estimate of the binary parameters. Improvements are
relatively smaller for lighter binaries.
More recently the Fisher matrix was employed to assess the errors
in estimating the parameters of nonspinning black-hole binaries
using the complete inspiral-merger-ring-down waveforms~\cite{AjithBose2009}.

Various authors have investigated the accuracy with which the LISA
detector can determine binary parameters including spin
effects. Cutler~\cite{cutler-98} determined LISA's angular resolution
and evaluated the errors of the binary masses and distance considering
spins aligned or anti-aligned with the orbital angular momentum.
Hughes~\cite{hughes-02} investigated the accuracy with which
the redshift can be estimated (if the cosmological parameters are
derived independently), and considered the black-hole ring-down phase
in addition to the inspiralling signal. Seto~\cite{seto-02} included
the effect of finite armlength (going beyond the long wavelength
approximation) and found that the accuracy of the distance
determination and angular resolution improve. This happens because the
response of the instrument when the armlength is finite depends
strongly on the location of the source, which is tightly correlated
with the distance and the direction of the orbital angular
momentum. Vecchio~\cite{vecchio-04} provided the first estimate of
parameters for precessing binaries when only one of the two
supermassive black holes carries spin. He showed that modulational
effects decorrelate the binary parameters to some extent, resulting in
a better estimation of the parameters compared to the case when spins
are aligned or antialigned with orbital angular momentum. Hughes and
Menou~\cite{hughes-05} studied a class of binaries, which they called
``golden binaries,'' for which the inspiral and ring-down phases could
be observed with good enough precision to carry out valuable tests of
strong-field gravity. Berti, Buonanno and Will~\cite{berti-05} have
shown that inclusion of non-precessing spin-orbit and spin-spin terms
in the gravitational-wave phasing generally reduces the accuracy with
which the parameters of the binary can be estimated. This is not
surprising, since the parameters are highly correlated, and adding
parameters effectively dilutes the available information.

Extensive study of accuracy of parameter estimation for continuous
gravitational-wave signals from spinning neutron stars was performed in~\cite{JaranowskiKrolak1999}.
In~\cite{Seto2005} Seto used the Fisher matrix
to study the possibility of determining distances
to rapidly rotating isolated neutron stars
by measuring the curvature of the wave fronts.

\subsubsection{Comparison with the Cram\`er--Rao bound}

In order to test the performance of the maximization method of the $\F$-statistic
it is useful to perform Monte Carlo simulations of the parameter estimation
and compare the simulated variances of the estimators
with the variances calculated from the Fisher matrix.
Such simulations were performed for various gravitational-wave signals~\cite{kokkotas-94,bala-95,jaranowski-00,Cokelaer2008}.
In these simulations one observes
that, above a certain signal-to-noise ratio, called the \emph{threshold signal-to-noise ratio},
the results of the Monte Carlo simulations agree very well
with the calculations of the rms errors from the inverse of the Fisher matrix.
However, below the threshold signal-to-noise ratio
they differ by a large factor.
This threshold effect is well known in signal processing~\cite{trees-68}.
There exist more refined theoretical bounds on the rms errors
that explain this effect, and they were studied
in the context of the gravitational-wave signals
from coalescing binaries~\cite{nicholson-98}.

Use of the Fisher matrix in the assessment of accuracy
of the parameter estimation has been critically examined in~\cite{vallisneri-07},
where a criterion has been established for the signal-to-noise ratio
above which the inverse of the Fisher matrix
approximates well covariances of the parameter estimators.
In~\cite{ZanolinVitaleMakris2010,VitaleZanolin2010}
the errors of ML estimators of parameters
of gravitational-wave signals from nonspinning black-hole binaries were calculated analytically
using a power expansion of the bias and the covariance matrix in inverse powers of the signal-to-noise ratio.
The first-order term in this covariance matrix expansion is the inverse of the Fisher information matrix.
The use of higher-order derivatives of the likelihood function in these expansions makes the errors prediction
sensitive to the secondary lobes of the pdf of the ML estimators.
Conditions for the validity of the Cram\`er--Rao lower bound are discussed in~\cite{VitaleZanolin2010} as well,
and some new features in regions of the parameter space so far not explored are predicted
(e.g., that the bias can become the most important contributor
to the parameters errors for high-mass systems with masses $200\,M_{\odot}$ and above).

There exists a simple model that explains
the deviations from the covariance matrix
and reproduces well the results of the Monte Carlo simulations
(see also~\cite{bala-98}).
The model makes use of the concept of the elementary cell
of the parameter space that we introduced in Section~\ref{sec:FA}.
The calculation given below is a generalization of the calculation of the
rms error for the case of a monochromatic signal given by Rife and
Boorstyn~\cite{rife-74}.

When the values of parameters of the template that correspond to the
maximum of the functional $\F$ fall within the cell in the parameter
space where the signal is present, the rms error is satisfactorily
approximated by the inverse of the Fisher matrix. However, sometimes,
as a result of noise, the global maximum is in the cell where there is
no signal. We then say that an \emph{outlier} has occurred. In the
simplest case we can assume that the probability density of the values
of the outliers is uniform over the search interval of a parameter,
and then the rms error is given by
\begin{equation}
  \sigma_\mathrm{out}^2 = \frac{\Delta^2}{12},
\end{equation}
where $\Delta$ is the length of the search interval for a given
parameter. The probability that an outlier occurs will be higher
the lower the signal-to-noise ratio is. Let $q$ be the probability
that an outlier occurs. Then the total variance $\sigma^2$ of the
estimator of a parameter is the weighted sum of the two errors
\begin{equation}
  \label{err}
  \sigma^2 = \sigma_\mathrm{out}^2 q + \sigma_\mathrm{CR}^2 (1 - q),
\end{equation}
where $\sigma_\mathrm{CR}$ is the rms errors calculated from the
covariance matrix for a given parameter. One can
show~\cite{jaranowski-00} that the probability $q$ can be approximated
by the following formula:
\begin{equation}
  \label{pout}
  q = 1 - \int^{\infty}_0 \!\!\! p_1(\rho,\F)
  \left(\int_0^\F \!\! p_0(y)\,
  \md y\right)^{\nc-1}\!\!\!\!\md\F,
\end{equation}
where $p_0$ and $p_1$ are the pdfs of the $\F$-statistic
(for known intrinsic parameters)
when the signal is absent or present in data, respectively
[they are given by Eqs.~(\ref{p0}) and~(\ref{p1})],
and where $\nc$ is the number of cells in the intrinsic parameter space.
Eq.~(\ref{pout}) is in good but not perfect
agreement with the rms errors obtained from the Monte Carlo
simulations (see~\cite{jaranowski-00}). There are clearly other
reasons for deviations from the Cram\`er--Rao bound as well. One important
effect (see~\cite{nicholson-98}) is that the functional $\F$ has many
local subsidiary maxima close to the global one. Thus, for a low
signal-to-noise ratio the noise may promote the subsidiary maximum to a
global one.


\subsection{Upper limits}

Detection of a signal is signified by a large value of the
$\F$-statistic that is unlikely to arise from the noise-only
distribution. If instead the value of $\F$ is consistent with pure
noise with high probability we can place an upper limit on the
strength of the signal. One way of doing this is to take the loudest
event obtained in the search and solve the equation
\begin{equation}
  \label{UL}
  P_\mathrm{D}(\rho_{\mathrm{UL}},\F_\mathrm{L}) = \beta
\end{equation}
for signal-to-noise ratio $\rho_\mathrm{UL}$, where $P_\mathrm{D}$ is
the detection probability given by Eq.~(\ref{PD}),
$\F_\mathrm{L}$ is the value of the $\F$-statistic corresponding to
the loudest event, and $\beta$ is a chosen confidence~\cite{astone-03,
  ligo3-04}. Then $\rho_{\mathrm{UL}}$ is the desired upper limit with
confidence $\beta$.

When gravitational-wave data do not conform to a Gaussian probability
density assumed in Eq.~(\ref{PD}), a more accurate upper limit
can be obtained by injecting the signals into the detector's data and
thereby estimating the probability of detection
$P_\mathrm{D}$~\cite{ligo1-04}.

\newpage

\section{Network of Detectors}
\label{sec:network-of-detectors}

Several gravitational-wave detectors can observe gravitational waves
from the same source. For example a network of bar detectors can
observe a gravitational-wave burst from the same supernova explosion,
or a network of laser interferometers can detect the inspiral of the
same compact binary system. The space-borne LISA detector can be
considered as a network of three detectors that can make three
independent measurements of the same gravitational-wave
signal. Simultaneous observations are also possible among different
types of detectors. For example, a search for supernova bursts can be
performed simultaneously by resonant and interferometric detectors~\cite{astone-94}.

Let us consider a network consisting of $N$ gravitational-wave detectors
and let us denote by $x_I$ the data collected by the $I$th detector ($I=1,\ldots,N$).
We assume that noises in all detectors are additive, so the data $x_I$
is a sum of the noise $n_I$ in the $I$th detector
and eventually a gravitational-wave signal $h_I$
registered by the $I$th detector,
\be
\label{r5.1}
x_I(t) = n_I(t) + h_I(t),
\quad I=1,\ldots,N.
\ee
It is convenient to collect all the data streams,
all the noises, and all the gravitational-wave signals
into column $N\times1$ matrices denoted respectively
by $\mathbf{x}$, $\mathbf{n}$, and $\mathbf{h}$,
\be
\label{r5.2}
\mathbf{x}(t) := \begin{pmatrix}
x_1(t)\\ \vdots \\ x_N(t) \end{pmatrix},\quad
\mathbf{n}(t) := \begin{pmatrix}
n_1(t)\\ \vdots \\ n_N(t) \end{pmatrix},\quad
\mathbf{h}(t) := \begin{pmatrix}
h_1(t)\\ \vdots \\ h_N(t) \end{pmatrix},
\ee
then Eqs.~\eqref{r5.1} can shortly be written as
\begin{equation}
\label{r5.3}
{\bf x}(t) = {\bf{n}}(t) + {\bf{h}}(t).
\end{equation}
If additionally all detectors' noises are stationary, Gaussian,
and continuous-in-time random processes with zero means,
the network log likelihood function is given by
\begin{equation}
\label{r5.4}
\log\Lambda[\mathbf{x}] = (\mathbf{x}|\mathbf{h})
- \frac{1}{2}(\mathbf{h}|\mathbf{h}),
\end{equation}
where the scalar product $(\,\cdot\,|\,\cdot\,)$
is defined by
\begin{equation}
\label{r5.5}
({\bf{x}}|{\bf{y}}):= 4\Re
\int^{\infty}_{0} \!\!
\tilde{\mathbf{x}}(f)^\mathsf{T} \cdot \mathbf{S}_n(f)^{-1}
\cdot \tilde{\bf{y}}(f)^{*}\,\md f.
\end{equation}
Here $\mathbf{S}_n$ is the \emph{one-sided} cross spectral density
matrix of the noises of the detector network, which is defined by (here
E denotes the expectation value)
\begin{equation}
\label{r5.6}
\mathrm{E}\left[\tilde{\mathbf{n}}(f)\cdot{\tilde{\mathbf{n}}^*(f')}^\mathsf{T}\right]
= \frac{1}{2} \, \delta(f-f') \, \mathbf{S}_n(|f|).
\end{equation}

The analysis is greatly simplified if the cross spectrum matrix
$\mathbf{S}_n$ is diagonal. This means that the noises in various detectors
are uncorrelated. This is the case when the detectors of the network
are in widely separated locations, like, for example, the two LIGO
detectors. However, this assumption is not always satisfied. An
important case is the LISA detector where the noises of the three
independent responses are correlated. Nevertheless for the case of
LISA one can find a set of three combinations for which the noises are
uncorrelated~\cite{prince-02,nayak-03}. When the cross spectrum
matrix is diagonal the network log likelihood function is just the sum of
the log likelihood functions for each detector.

Derivation of the likelihood function
for an arbitrary network of detectors can be found in~\cite{finn-01}.
Applications of optimal filtering
for observations of gravitational-wave signals from coalescing binaries
by networks of ground-based detectors are given
in~\cite{jaranowski-94,cutler-94,JaranowskiOthers1996,bose-01},
and for the case of stellar-mass binaries observed by LISA space-borne detector
in~\cite{krolak-04,Rogan2004}.
The single-detector $\F$-statistic for nearly monochromatic gravitational waves
from spinning neutron stars was generalized to the case of a network of detectors
(also with time-varying noise curves) in~\cite{cutler-05} (in this work
the $\F$-statistic was also generalized from the usual single-source case
to the case of a collection of known sources).
The reduced Fisher matrix [defined in Eq.~\eqref{eq:rF}]
for the case of a network of interferometers observing spinning neutron stars
has been derived and studied in~\cite{prix-07}.

Network searches for gravitational-wave burst signals of unknown shape
are often based on maximization of the network likelihood function
over each sample of the unknown polarization waveforms $h_+$ and $h_\times$
and over sky positions of the source~\cite{FlanaganHughes1998,MohantyOthers2004}.
A least-squares-fit solution for the estimation of
the sky location of the source and the polarization waveforms
by a network of three detectors for the case of a broadband burst was obtained
in~\cite{gursel-89}.

There is also another important method for analyzing the data from a
network of detectors -- the search for coincidences of events among
detectors. This analysis is particularly important when we search for
supernova bursts, the waveforms of which are not very well known. Such
signals can be easily mimicked by the non-Gaussian behavior of the
detector noise. The idea is to filter the data optimally in each of
the detectors and obtain candidate events. Then one compares parameters
of candidate events, like, for example, times of arrivals of the bursts,
among the detectors in the network. This method is widely used in the
search for supernovae by networks of bar detectors~\cite{astone2-03}.
A new geometric coincident algorithm of combining the data from a network of detectors
was proposed in~\cite{Robinson_et_al2008}.
This algorithm employs the covariances between signal's parameters
in such a way that it associates with each candidate event
an ellipsoidal region in parameter space
defined by the covariance matrix.
Events from different detectors are deemed to be in coincidence
if their ellipsoids have a nonzero overlap.
The coincidence and the coherent strategies of multidetector detection
of gravitational-wave signals from inspiralling compact binaries have been
compared in~\cite{Mukhopadhyay_et_al2006,Mukhopadhyay_et_al2009,Bose2011}.
~\cite{Mukhopadhyay_et_al2006} considered detectors in pairs located in the same site
and~\cite{Mukhopadhyay_et_al2009} pairs of detectors at geographically separated sites.
The case of three detectors (like the network of two LIGO detectors and the Virgo detector)
has been considered in detail in ~\cite{Bose2011}, where it was demonstrated
that the hierarchical coherent pipeline on Gaussian data has a better performance
than the pipeline with just the coincident stage.

A general framework for studying the effectiveness of networks of
interferometric gravitational-wave detectors has been proposed
in~\cite{Schutz2011}. Using this framework it was shown that adding a
fourth detector to the existing network of LIGO/VIRGO detectors can
dramatically increase, by a factor of 2 to 4, the detected event rate
by allowing coherent data analysis to reduce the spurious instrumental
coincident background.

\newpage

\section{Non-stationary, Non-Gaussian, and Non-linear Data}
\label{sec:non}

Eqs.~(\ref{eq:aMLE}) and (\ref{eq:FS}) provide maximum likelihood
estimators only when the noise in which the signal is buried is
Gaussian. There are general theorems in statistics indicating that the
Gaussian noise is ubiquitous. One is the \emph{central limit theorem},
which states that the mean of any set of variables with any
distribution having a finite mean and variance tends to the normal
distribution. The other comes from the information theory and says
that the probability distribution of a random variable with a given
mean and variance, which has the maximum entropy (minimum information)
is the Gaussian distribution. Nevertheless, analysis of the data from
gravitational-wave detectors shows that the noise in the detector may
be non-Gaussian (see, e.g., Figure~6 in~\cite{astone-93}). The noise
in the detector may also be a non-linear and a non-stationary random
process.

The maximum likelihood method does not require that the noise in the
detector be Gaussian or stationary. However, in order to derive the
optimum statistic and calculate the Fisher matrix we need to know the
statistical properties of the data. The probability distribution of
the data may be complicated, and the derivation of the optimum
statistic, the calculation of the Fisher matrix components and the
false alarm probabilities may be impractical. However, there is one
important result that we have already mentioned. The matched-filter,
which is optimal for the Gaussian case is also a linear filter that
gives maximum signal-to-noise ratio no matter what the distribution
of the data. Monte Carlo simulations performed by Finn~\cite{finn-01}
for the case of a network of detectors indicate that the performance
of matched-filtering (i.e., the maximum likelihood method for Gaussian
noise) is satisfactory for the case of non-Gaussian and stationary
noise.

Allen et~al.~\cite{allenetal-02, allenetal-03} derived an optimal
(in the Neyman--Pearson sense, for weak signals)
signal processing strategy, when the detector noise is
non-Gaussian and exhibits tail terms.
This strategy is robust, meaning that it is close to optimal for Gaussian noise
but far less sensitive than conventional methods to the excess large events
that form the tail of the distribution.
This strategy is based on a \emph{locally optimal test}~\cite{kassam-88}
that amounts to comparing a first non-zero derivative
\begin{equation}
\Lambda_n[x] = \frac{\md^n \Lambda[x|\epsilon]}{\md\epsilon^n}\bigg|_{\epsilon = 0}
\end{equation}
of the likelihood ratio with respect to the amplitude of the signal
with a threshold instead of the likelihood ratio itself.

The non-stationarity in the case of Gaussian and uncorrelated noise can be easily incorporated
into matched filtering (see Appendix~C of~\cite{ligovirgo1-11}).
Let us assume that a noise sample $n_l$ in the data has a Gaussian pdf
with a variance $\sigma^2_l$ and zero mean ($l=1,\ldots,N$, where $N$ is the number of data points).
Different noise samples may have distributions with different variances.
We also assume that the noise samples are uncorrelated,
then the autocorrelation function $K(l,l')$ of the noise is given by
[see Eq.~\eqref{Kn}]
\begin{equation}
\label{eq:auto}
K(l,l') = \sigma^2_l\,\delta_{ll'},
\end{equation}
where $\delta_{ll'}$ is the Kronecker delta function.
In the case of a known signal $h_l$ and additive noise
the optimal filter $q_l$ is the solution of the following equation
[which is a discrete version of the integral Eq.~\eqref{IntEq}]:
\begin{equation}
\label{eq:IntEqWN}
h_l = \sum^N_{l'=1} K(l,l')\,q_{l'}.
\end{equation}
Thus, we have the following equation for the filter $q_l$:
\begin{equation}
q_l = \frac{h_l}{\sigma^2_l},
\end{equation}
and the following expression for the log likelihood ratio:
\begin{equation}
\ln\Lambda[x] = \sum^N_{l=1} \frac{h_l x_l}{\sigma^2_l}
- \frac{1}{2} \sum^N_{l=1} \frac{h_l^2}{\sigma^2_l}.
\end{equation}
Thus, we see that for non-stationary, uncorrelated Gaussian noise
the optimal processing is identical to matched filtering for a known signal in stationary Gaussian noise,
except that we divide both the data $x_l$ and the signal $h_l$ by time-varying standard deviation of the noise.
This may be thought of as a special case of whitening the data and then correlating it using a whitened filter.

In the remaining part of this section we review some statistical tests
and methods to detect non-Gaussianity, non-stationarity, and
non-linearity in the data. A classical test for a sequence of data to
be Gaussian is the Kolmogorov--Smirnov test~\cite{conover-80}. It
calculates the maximum distance between the cumulative distribution of
the data and that of a normal distribution, and assesses the
significance of the distance. A similar test is the Lillifors
test~\cite{conover-80}, but it adjusts for the fact that the
parameters of the normal distribution are estimated from the data
rather than specified in advance. Another test is the Jarque--Bera
test~\cite{judge-80}, which determines whether sample skewness and
kurtosis are unusually different from their Gaussian values.

A useful test to detect outliers in the data is Grubbs' test~\cite{Grubbs1969}.
This test assumes that the data has an underlying Gaussian probability distribution
but it is corrupted by some disturbances.
Grubbs' test detects outliers iteratively. Outliers are removed one by one
and the test is iterated until no outliers are detected.
Grubbs' test is a test of the null hypothesis:
\begin{enumerate}
\item $\mathrm{H}_0$: \emph{There are no outliers in the data set $x_k$.}
\suspend{enumerate}
against the alternate hypothesis:
%
\resume{enumerate}
\item $\mathrm{H}_1$: \emph{There is at least one outlier in the data set $x_k$.}
\end{enumerate}
The Grubbs' test statistic is the largest absolute deviation from the sample mean
in units of the sample standard deviation, so it is defined as
\begin{equation}
G = \frac{\max_k|x_k-\mu|}{\sigma},
\end{equation}
where $\mu$ and $\sigma$ denote the sample mean and the sample standard deviation, respectively.
The hypothesis of no outliers is rejected if
\begin{equation}
G > \frac{n - 1}{\sqrt{n}}\sqrt{\frac{t^2_{\alpha/(2 n), n - 2}}{n - 2 + t^2_{\alpha/(2 n), n - 2}}},
\end{equation}
where $t_{\alpha/(2 n), n - 2}$  denotes the critical value of the t-distribution
with $n-2$ degrees of freedom and a significance level of $\alpha/(2 n)$.

Grubbs' test has been used to identify outliers in the search of Virgo data
for gravitational-wave signals from the Vela pulsar~\cite{ligovirgo1-11}.
A test to discriminate spurious events due to non-stationarity and non-Gaussianity of the data
from genuine gravitational-wave signals has been developed by Allen~\cite{Allen2005}.
This test, called the \emph{$\chi^2$ time-frequency discriminator},
is applicable to the case of broadband signals, such as those coming from compact coalescing binaries.

Let now $x_k$ and $u_l$ be two discrete-in-time random processes
($-\infty<k,l<\infty$) and let $u_l$ be independent and identically
distributed (i.i.d.) random variables.
We call the process $x_k$ \emph{linear} if it can be represented by
\begin{equation}
  \label{lp}
  x_k = \sum_{l=0}^{N} a_l u_{k-l},
\end{equation}
where $a_l$ are constant coefficients. If $u_l$ is Gaussian
(non-Gaussian), we say that $x_l$ is linear Gaussian
(non-Gaussian). In order to test for linearity and Gaussianity we
examine the third-order cumulants of the data. The third-order
cumulant $C_{kl}$ of a zero mean stationary process is defined by
\begin{equation}
  \label{cum}
  C_{kl} := \mathrm{E} \left[ x_m x_{m + k} x_{m + l} \right].
\end{equation}
The bispectrum $S_2(f_1, f_2)$ is the two-dimensional Fourier
transform of $C_{kl}$. The bicoherence is defined as
\begin{equation}
  \label{bis}
  B(f_1, f_2) := \frac{S_2(f_1, f_2)}{S(f_1 + f_2) S(f_1) S(f_2)},
\end{equation}
where $S(f)$ is the spectral density of the process $x_k$. If the
process is Gaussian, then its bispectrum and consequently its
bicoherence is zero. One can easily show that if the process is linear
then its bicoherence is constant. Thus, if the bispectrum is not zero,
then the process is non-Gaussian; if the bicoherence is not constant
then the process is also non-linear. Consequently we have the
following hypothesis testing problems:
\begin{enumerate}
\item $\mathrm{H}_1$: \emph{The bispectrum of $x_k$ is nonzero}.
  \label{item_1}
\item $\mathrm{H}_0$: \emph{The bispectrum of $x_k$ is zero}.
  \label{item_2}
\suspend{enumerate}
If Hypothesis~\ref{item_1} holds, we can test for linearity, that is, we
have a second hypothesis testing problem:
\resume{enumerate}
%
\item $\mathrm{H}_1'$: \emph{The bicoherence of $x_k$ is not constant}.
  \label{item_3}
\item $\mathrm{H}_1''$: \emph{The bicoherence of $x_k$ is a constant}.
  \label{item_4}
\end{enumerate}
If Hypothesis~\ref{item_4} holds, the process is linear.

Using the above tests we can \emph{detect} non-Gaussianity and, if the
process is non-Gaussian, non-linearity of the process. The
distribution of the test statistic $B(f_1, f_2)$,
Eq.~(\ref{bis}), can be calculated in terms of $\chi^2$
distributions. For more details see~\cite{hinich-82}.

It is not difficult to examine non-stationarity of the data. One can
divide the data into short segments and for each segment calculate the
mean, standard deviation and estimate the spectrum. One can then
investigate the variation of these quantities from one segment of the
data to the other. This simple analysis can be useful in identifying
and eliminating bad data. Another quantity to examine is the
autocorrelation function of the data. For a stationary process the
autocorrelation function should decay to zero. A test to detect
certain non-stationarities used for analysis of econometric time
series is the Dickey--Fuller test~\cite{brooks-02}. It models the data
by an autoregressive process and it tests whether values of the
parameters of the process deviate from those allowed by a stationary
model. A robust test for detecting non-stationarity in data from
gravitational-wave detectors has been developed by
Mohanty~\cite{mohanty-00}. The test involves applying Student's t-test
to Fourier coefficients of segments of the data.
Still another \emph{block-normal} approach
has been studied by McNabb et al.~\cite{McNabb2004}.
It identifies places in the data stream where the characteristic statistics of the data change.
These change points divide the data into blocks in  characteristics are stationary.

\newpage


\section{Acknowledgments}

One of us (AK) acknowledges support from the National Research
Council under the Resident Research Associateship program at the
Jet Propulsion Laboratory, California Institute of Technology and
from Max-Planck-Institut f\"ur Gravitationsphysik.
This research was supported in part by the MNiSzW grants nos.\ 1 P03B
029 27 and N N203 387237.

\newpage


\bibliography{refs}

\end{document}